\documentclass[reprint,superscriptaddress,preprintnumbers,amsmath,amssymb,aps,prd,tightenlines,longbibliography,nofootinbib,balancelastpage]{revtex4-2}

\usepackage{graphicx}
\usepackage[dvipsnames]{xcolor}
\usepackage[sort&compress]{natbib}
\usepackage{amsmath,amssymb,bm,bbm,slashed,subdepth}
\usepackage{xr-hyper}
\usepackage[colorlinks=true
,urlcolor=blue
,anchorcolor=blue
,citecolor=blue
,filecolor=blue
,linkcolor=red
,menucolor=blue
,linktocpage=true
,pdfproducer=medialab
,pdfa=true
]{hyperref}
\usepackage{cleveref}
\usepackage{enumerate}
\usepackage{epsfig, subfigure}
\usepackage{setspace}
\usepackage{booktabs, tabularx}
\usepackage{units}
\usepackage{placeins}
\usepackage{multirow}
\usepackage{mathtools}
\usepackage[normalem]{ulem}
\usepackage{esint}
\usepackage{float}
\usepackage{comment}
\usepackage{braket}
\usepackage{textgreek}

\usepackage{silence}
\newcommand{\be}{\begin{eqnarray}}
\newcommand{\ee}{\end{eqnarray}}

\newcommand{\sN}{{\scriptscriptstyle N}}
\newcommand{\sEM}{{\scriptscriptstyle {\rm EM}}}
\newcommand{\DM}{{\scriptscriptstyle {\rm DM}}}
\newcommand{\SP}{{\scriptscriptstyle {\rm SP}}}
\newcommand{\QM}{{\scriptscriptstyle {\rm QM}}}
\newcommand{\bx}{\mathbf{x}}
\newcommand{\bn}{\mathbf{n}}
\newcommand{\bB}{\mathbf{B}}
\newcommand{\bH}{\mathbf{H}}
\newcommand{\bS}{\mathbf{S}}
\newcommand{\bM}{\mathbf{M}}
\newcommand{\bA}{\mathbf{A}}
\newcommand{\bE}{\mathbf{E}}
\newcommand{\bJ}{\mathbf{J}}
\newcommand{\hx}{\hat{\mathbf{x}}}
\newcommand{\hy}{\hat{\mathbf{y}}}
\newcommand{\hz}{\hat{\mathbf{z}}}
\newcommand{\bmu}{\boldsymbol \mu}
\newcommand{\gagg}{g_{a\gamma\gamma}}
\newcommand{\w}{\omega}
\newcommand{\pd}{\partial}

\definecolor{light-gray}{gray}{0.9}

\begin{document}

\title{Dark Matter Nuclear Magnetic Resonance is Sensitive\\to Dark Photons and the Axion-Photon Coupling}

\author{Carl~Beadle}
\affiliation{D\'epartement de Physique Th\'eorique, Universit\'e de Gen\`eve, 
24 quai Ernest Ansermet, 1211 Gen\`eve 4, Switzerland}

\author{Sebastian~A.~R.~Ellis}
\affiliation{D\'epartement de Physique Th\'eorique, Universit\'e de Gen\`eve, 
24 quai Ernest Ansermet, 1211 Gen\`eve 4, Switzerland}

\author{Jacob~M.~Leedom}
\affiliation{CEICO, Institute of Physics of the Czech Academy of Sciences,\\
Na Slovance 2, 182 00 Prague 8, Czech Republic}

\author{Nicholas~L.~Rodd}
\affiliation{Theory Group, Lawrence Berkeley National Laboratory, Berkeley, CA 94720, USA}
\affiliation{Berkeley Center for Theoretical Physics, University of California, Berkeley, CA 94720, USA}

\begin{abstract}
We demonstrate that nuclear magnetic resonance based searches for dark matter (DM) have intrinsic and powerful sensitivity to dark photons and the axion-photon coupling.
The reason is conceptually straightforward.
An instrument such as CASPEr-Gradient begins with a large sample of nuclear spins polarised in a background magnetic field.
In the presence of axion DM coupled to nucleons, the spin ensemble feels an effective magnetic field $\bB \propto \nabla a$ that tilts the spins, generating a potentially observable precession.
If the magnetic field is real rather than effective, the system responds identically.
A real field can be generated by a kinetically mixed dark photon within the shielded region the sample is placed or an axion coupled to photons through its interaction with the background magnetic field.
We show that all three signals are detectable and distinguishable. 
If CASPEr-Gradient were to reach the QCD axion prediction of the axion-nucleon coupling, it would simultaneously be sensitive to kinetic mixings of $\epsilon \simeq 3 \times 10^{-16}$ and axion-photon couplings of $\gagg \simeq 2 \times 10^{-16}\,{\rm GeV}^{-1}$ for $m \simeq 1\,\mu{\rm eV}$.
\end{abstract}

\maketitle

Particle dark matter (DM) remains the most compelling hypothesis to explain the universe's missing matter.
However, evidence for its existence does not pinpoint a particular mass or strength of coupling to visible matter.
Viable particle DM parameter space therefore spans a daunting range of masses.
Historically, WIMPs and the electroweak scale have taken preeminence.
More recently the community has cast a broader net, exemplified by the efforts to search for ultralight DM ($m \ll 10\,{\rm eV}$), where DM transitions to a classical wave description~\cite{Cheong:2024ose}.

A question less often considered is what is the full range of physics that our nets may catch.
Planned experiments aim to cover enormous swathes of unexplored ultralight DM parameter space.
Unexpected signals could emerge.
For example, these devices can detect relativistic signals such as a cosmic axion background~\cite{Dror:2021nyr,ADMX:2023rsk} or gravitational waves~\cite{Berlin:2021txa,Berlin:2023grv,Domcke:2022rgu,Domcke:2023bat,Domcke:2024mfu,Pappas:2025zld,TitoDAgnolo:2024uku}.
With this Letter we instead demonstrate that \textit{even DM signals have been overlooked}: devices designed to search for one type of wave DM can be strongly sensitive to another.
More to the point, ongoing efforts to detect axion DM coupled to nucleons with nuclear magnetic resonance (NMR), for example by CASPEr-Gradient~\cite{Graham:2013gfa,Budker:2013hfa}, would be highly sensitive to the axion-photon coupling and dark photon DM, as is the case in other torque based searches~\cite{Kalia:2024eml}.
Given our ignorance as to the nature of DM, it stands that these situations should be highlighted whenever possible.

To justify our claim we recall the basics of axion NMR.
Consider a sample of nuclear spins aligned in a background magnetic field $\bB_0$.
In conventional NMR, one fires an electromagnetic wave through the sample that generates an interaction according to ${\cal H}_B = - \gamma \bB \cdot \bS$, with $\gamma$ and $\bS$ the nucleon gyromagnetic ratio and spin operator.
Consequently, the spins are tipped into the plane transverse to $\bB_0$ and begin to precess at the Larmor frequency, $\w_0 = \gamma B_0$, yielding an oscillatory signal in the transverse plane.
CASPEr-Gradient exploits the same physical principle.
If the spins are immersed in an axion DM wave, then the axion-nucleon coupling, ${\cal L}_{a\sN} = g_{\sN} (\pd_{\mu} a) \bar{N} \gamma^{\mu} \gamma_5 N$, generates an interaction ${\cal H}_{a\sN} = - 2 g_{\sN} \nabla a \cdot \bS$.
By analogy to ${\cal H}_B$ the axion induces an \textit{effective} magnetic field of the form $\bB_{a\sN} = g_{\sN}(2/\gamma) \nabla a$.
The axion field has an identical effect: DM tips the spins, which then precess leading to a detectable signal.
There is a resonant enhancement when $\w_0 \simeq m$ and therefore one performs a resonant search by varying $B_0$.

\begin{figure*}[!t]
\centering
\includegraphics[width=0.45\textwidth]{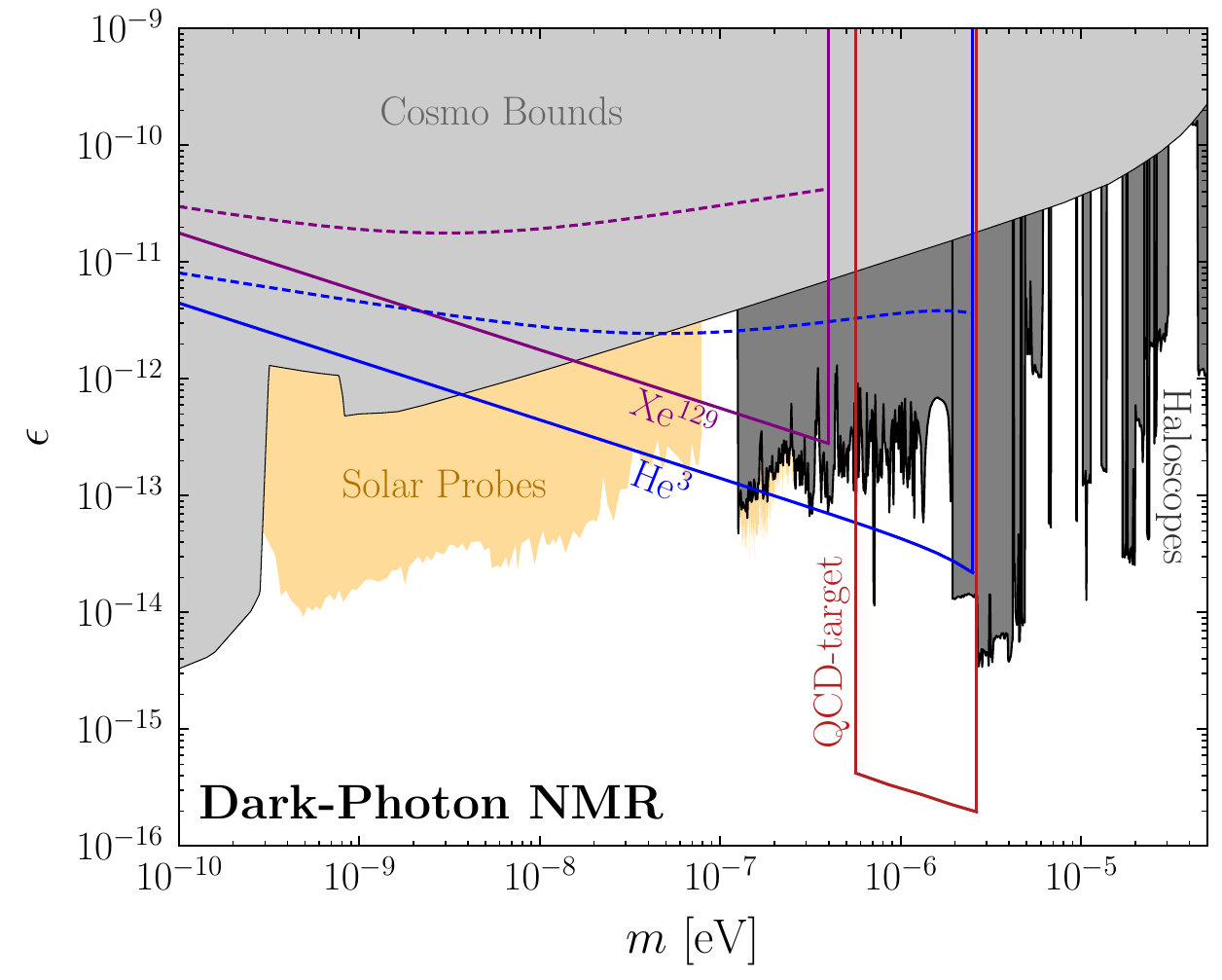}
\hspace{0.5cm}
\includegraphics[width=0.45\textwidth]{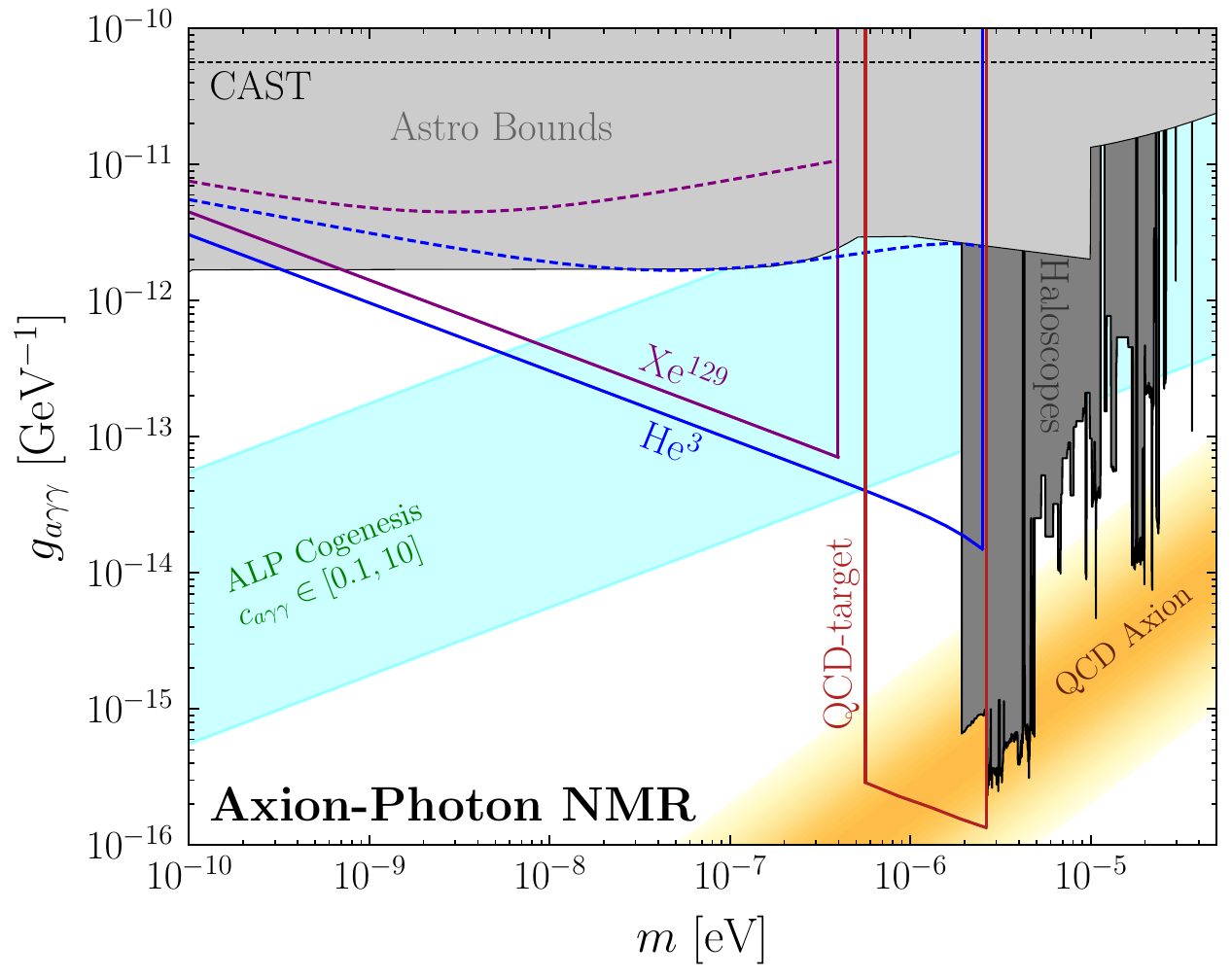}
\vspace{-0.4cm}
\caption{Projected reach of DM NMR to dark photons (left) and axions (right).
Both searches can be performed \textit{simultaneously} along with a search for the axion nucleon coupling.
The purple and blue lines correspond to a sample constituted of xenon-129 ($\gamma = 12$\,MHz/T, $n = 1.3 \times 10^{22}$\,cm$^{-3}$, $H_{\text{max}} = 10$\,T) and helium-3 ($\gamma = 32$\,MHz/T, $n = 2.8 \times 10^{22}$\,cm$^{-3}$, $H_{\text{max}} = 20$\,T).
Solid curves assume $T_2=100$\,s, whereas dashed curves assume a line width limited by a part-per-million.
A search capable of reaching the QCD axion prediction for $g_{\sN}$ would have the sensitivity shown in red~\cite{Dror:2022xpi}.
The reach assumes a single TM mode with the sample placed at the optimal radius of $r \simeq 0.77 R$.
See the text for further details.
}
\vspace{-0.5cm}
\label{fig:Results}
\end{figure*}

We can now explain our core insight: if DM generates a real rather than effective magnetic field, CASPEr-Gradient can detect it.
Consider first dark photon DM, $\bA'$, that mixes with the visible photon via a kinetic mixing parameter $\epsilon$ \cite{Holdom:1985ag}.
To mitigate the impact of stray magnetic fields, the spin sample must be placed within a shielded region of geometric size $L$.
The dark photon will penetrate that shield and generate a magnetic field $\bB_{A'} \sim \epsilon m^2 L \bA'$~\cite{Chaudhuri:2014dla}.
Setting $B_{aN} \sim B_{A'}$, the qualitative dark-photon sensitivity in terms of the axion-nucleon coupling is $\epsilon \sim g_{\sN} v / \gamma mL$, with $v \sim 10^{-3}$ the local DM speed.
We can rephrase the sensitivity as follows.
Writing $g_{\sN} = c_{\sN} m/m_\pi f_\pi$, which depends on the pion mass and decay constant, we have
\be
\epsilon \sim 10^{-16}\, c_{\sN} \left( \gamma_{\rm He}/\gamma \right) \left( 10\,{\rm cm}/L \right)\!,
\label{eq:eps-qual}
\ee
assuming a helium-3 sample.
Although an ambitious target, the ultimate goal of all axion experiments is QCD sensitivity, where $c_{\sN} \sim 1$.
Note the above scaling is distinct from ideas to repurpose axion-photon searches for the dark photon, as in e.g. Ref.~\cite{Caputo:2021eaa}.
In those cases the dark photon sensitivity does not gain from the large magnetic field volume that is maximised for the axion sensitivity.
Here, however, the scaling of both the axion and dark photon signal is controlled by the density of spins, or effectively the background magnetization.

A similar argument applies to the axion-photon coupling, ${\cal L}_{a\gamma\gamma} = -\tfrac{1}{4} g_{a\gamma\gamma} a F\widetilde{F}$, where axion DM couples to $\bB_0$ and induces $B_{a\gamma} \sim \gagg B_0 L (\pd_t a)$, where now $L$ is the scale over which $B_0$ is generated.
The expected sensitivity from $B_{aN} \sim B_{a\gamma}$ is $\gagg \sim g_N v/\gamma B_0 L$, or taking $\gagg \sim c_{\gamma} \alpha_{\sEM} m/2\pi m_\pi f_\pi$,
\be
c_{\gamma} \sim c_{\sN} \left( 10\,{\rm T}/ B_0 \right) \left( \gamma_{\rm He} / \gamma \right) \left( 10\,{\rm cm} /L \right)\!.
\label{eq:cg-qual}
\ee
This result demonstrates that at the highest frequencies -- recall $B_0$ is tuned to adjust the resonant frequency -- an instrument sensitive to the QCD axion value of $c_{\sN}$ would also be sensitive to the QCD axion prediction of $c_{\gamma} \sim 1$.

In what follows we refine the qualitative results of Eqs.~\eqref{eq:eps-qual} and \eqref{eq:cg-qual} culminating in the projections shown in Fig.~\ref{fig:Results}.
These results show the reach for several sensitivity estimates for CASPEr-Gradient that exist in the literature, specifically Refs.~\cite{JacksonKimball:2017elr,Dror:2022xpi}, each shown for two different target nuclei, xenon-129 and helium-3.
The results are compared to existing constraints from Refs.~\cite{An:2024wmc,An:2023wij,Paola_Arias_2012,Caputo_2020,McDermott_2020,PhysRevLett.105.171801,Nguyen_2019,PhysRevLett.124.101802,Zhong_2018,Backes_2021,PhysRevD.104.012013,ahn2024extensivesearchaxiondark,Alesini:2020vny,Ahyoune:2024klt,Quiskamp_2022,Anastassopoulos2017,PhysRevLett.131.111004,PhysRevD.105.103034,PhysRevD.107.083027,ABE2024101425,PhysRevLett.116.161101,Benabou:2025jcv} and the  theoretical predictions of Refs.~\cite{Dine:1981rt,PhysRevLett.43.103,Shifman:1979if,Co_2021}, see Ref.~\cite{AxionLimits}.

\vspace{0.2cm}
\noindent {\bf Principles of DM NMR.}
%
In the presence of a magnetic field, a spin with magnetic moment $\bmu = \gamma \bS$ experiences a torque, and therefore precesses according to $\dot{\bmu} = \bmu \times \gamma \bB$.
Taking $\bB = B_0 \hz$, the transverse coordinates $\mu_{\pm} = (\mu_x \pm i\mu_y)/\sqrt{2}$ evolve under the precession as $\mu_{\pm}(t) \propto e^{\mp i \w_0 t}$.
NMR extends this picture to a macroscopic sample of spins: $\bmu$ is promoted to $\bM$, the magnetization within the material, and $\bB$ to $\bH$.
We assume $\bH = H_0 \hz$ up to small corrections, however, the dynamics in the sample is also controlled by spin-spin and spin-lattice interactions.
These interactions exponentially damp coherent oscillations in the transverse and longitudinal directions, respectively, and are quantified by corresponding timescales $T_2$ and $T_1$.
The system evolves according to the Bloch equations,
\be
\dot{\bM} = \bM \times \gamma \bH - \frac{M_x \hx+M_y \hy}{T_2} - \frac{(M_z-M_0)\hz}{T_1},
\label{eq:bloch}
\ee
with $M_0$ the steady state magnetization generated by $H_0$.

We now introduce DM into the system as a small correction to the magnetic field, $\bH = H_0 \hz + \bH_\DM$.
While the aim of CASPEr-Gradient is to detect the effective magnetic field $\bH_\DM^{\rm eff} = g_{\sN} (2/\gamma) \nabla a$, the goal of this paper is to highlight a broader range of possibilities.
Taking $H_\DM/H_0 \ll 1$, we solve the problem perturbatively.
At zeroth order, the steady state solution is $\bM = M_0 \hz$.
In general, both transverse components of $\bM$ and $\bH_{\DM}$ are coupled and must be accounted for carefully.
If, however, we define $M_{\pm} = (M_x \pm i M_y)/\sqrt{2}$ and similarly $H_{\DM}^{\pm}$ the first order Bloch equations decouple, leaving
\be
\dot{M}_+ = - T_2^{-1} M_+ - i (\w_0 M_+ - \gamma M_0 H_\DM^+).
\label{eq:Bloch-pm}
\ee
Taking $M_+(0)=0$ the solution to this equation is
\be
\hspace{-0.5cm}M_+(t) = i \gamma M_0 \int_0^t dt'\, e^{-(t-t')/T_2} e^{-i \w_0(t-t')} H_\DM^+(t').
\label{eq:Mpsol}
\ee
Accordingly, for a perfectly resonant DM field, $H_\DM^+ = H_\DM e^{-i \w_0 t}$, the magnetization grows as
\be
M_+(t) = i e^{-i \w_0 t}\, \gamma M_0 H_\DM T_2\, (1-e^{-t/T_2}).
\label{eq:Mpres}
\ee
For $t \ll T_2$ DM drives a linear growth in the transverse component of the magnetization which continues until the amplitude begins to saturate once $t \sim T_2$.
The dynamics are comparable for a more realistic model of a DM field with finite coherence.
An exception occurs if the DM coherence time, $\tau_\DM$, is shorter than $T_2$, in which case for $\tau_\DM < t < T_2$ the growth of $M_+$ slows from linear to $\sqrt{t}$~\cite{Dror:2022xpi}.
Regardless, for $t \gg T_2,\tau_\DM$ the amplitude of the transverse magnetization saturates at $\gamma M_0 H_\DM T_2$.

In summary, DM drives the growth of a transverse oscillatory component to the magnetization.
To provide a sensitivity estimate we must determine the appropriate backgrounds.
Firstly, the spins within the sample generate an intrinsic noise source according to the fluctuation dissipation theorem.
Specifically, Eq.~\eqref{eq:Mpsol} defines a susceptibility of $\chi_{++}(t) = i \gamma M_0 e^{-t/T_2} e^{-i \w_0 t} \Theta(t)$ or $\chi_{++}(\w) = i \gamma M_0/[i(\w_0-\w)+T_2^{-1}]$.
The imaginary part of this expression dictates the power spectral density (PSD) of the noise,
\be
S^{\SP}_{M_+}(\w) \simeq \frac{2\gamma M_0 T_s}{\w V} \frac{T_2}{1+(\w-\w_0)^2 T_2^2}.
\label{eq:SPNoise}
\ee
This background is none other than the spin projection noise.
The noise depends on the bulk magnetization $M_0 = n \gamma \w_0/4T_s$, with $n$ the spin density in the sample and $T_s$ the spin temperature---from this we confirm the usual result that the form is independent of temperature.

The combined effect of DM and spin-projection noise is a transverse magnetization that leads to a corresponding magnetic field outside the sample.
If the magnetization is measured inductively through a SQUID, there is an additional background contribution from the magnetometer itself.
Further, thermal noise sources throughout the device contribute additional backgrounds.
Nevertheless, for the parameters CASPEr intends to employ the dominant effect is spin projection noise.

Accordingly, we can estimate the sensitivity by searching for a DM signal over the spin projection background.
Working at high masses where $T_2 > \tau_{\DM}$, we imagine interrogating the system for a time $T=T_2$ so that the observed power is distributed over a frequency bin of width $2\pi/T_2$.
For the bin centred around $\w_0$, the average spin projection noise power is $\bar{P}_{\SP} = \gamma^2 n T_2/16\pi V$.
For the signal, after accounting for the finite coherence, the equivalent expression is $\bar{P}_{\DM} = (\gamma M_0 H_{\DM} T_2)^2 \tau_a/2\pi$.
Assuming the signal and background magnetizations undergo Gaussian fluctuations, the power is distributed exponentially and we therefore expect the sensitivity limit to occur for $\bar{P}_{\DM} = \bar{P}_{\SP}$, or neglecting ${\cal O}(1)$ factors
\be
H_{\DM} \sim \frac{v}{\gamma} \sqrt{\frac{m}{nT_2V}},
\label{eq:HDM-sens}
\ee
where we took $M_0 = n (\gamma/2)$, assuming a hyperpolarised sample, and $\tau_a \sim 1/m v^2$.
A longer measurement with $T > T_2$ would improve on this sensitivity by $(T/T_2)^{1/4}$; further, when $T_2 > \tau_a$ a gain can be made by accounting for the multiple frequency bins in which the signal appears~\cite{Dror:2022xpi}.
For the axion-nucleon coupling, where $H_{\DM} = g_{\sN} (2/\gamma) \nabla a \sim g_{\sN} (v/\gamma) \sqrt{\rho_{\DM}}$, Eq.~\eqref{eq:HDM-sens} gives an estimated sensitivity of $g_{\sN} \sim \sqrt{m/\rho_{\DM} n T_2 V}$.

This ends our compact review of DM NMR.
A more detailed discussion appears in the Supplemental Material (SM) and we now turn to the different forms of $\bH_\DM$.

\vspace{0.2cm}
\noindent {\bf Dark Photon NMR.}
%
Consider the scenario where DM is a boson of a dark gauged U(1): a dark photon $A^{\prime \mu}$.
(We leave our discussion UV agnostic and refer instead to Refs.~\cite{Cyncynates:2024yxm,East:2022rsi}.)
This state can couple to the visible sector through the marginal operator $\tfrac{1}{2} \epsilon F^{\mu \nu} F'_{\mu \nu}$ parameterised by a kinetic mixing $\epsilon$ taken to be small throughout.
Rotating the two gauge fields to remove the kinetic mixing transforms to the interaction basis, where the dark photon remains decoupled from standard model currents yet provides a source for the visible U(1), quantified by $J_{\rm eff}^{\mu} \simeq - \epsilon m^2 A^{\prime \mu}$.

\begin{figure*}[!t]
\centering
\includegraphics[height=0.273\textwidth]{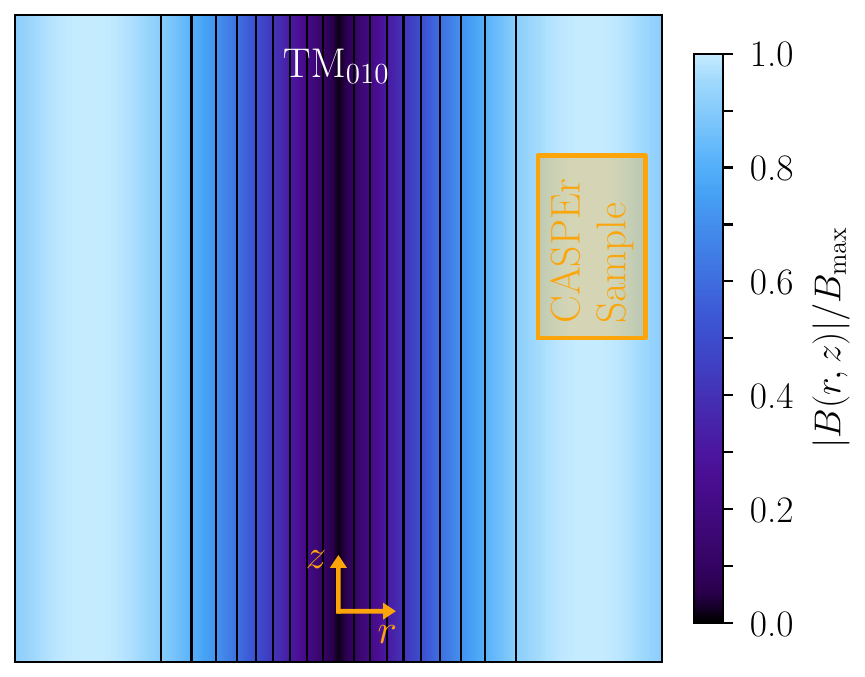}\hspace{0.1cm}
\includegraphics[height=0.273\textwidth]{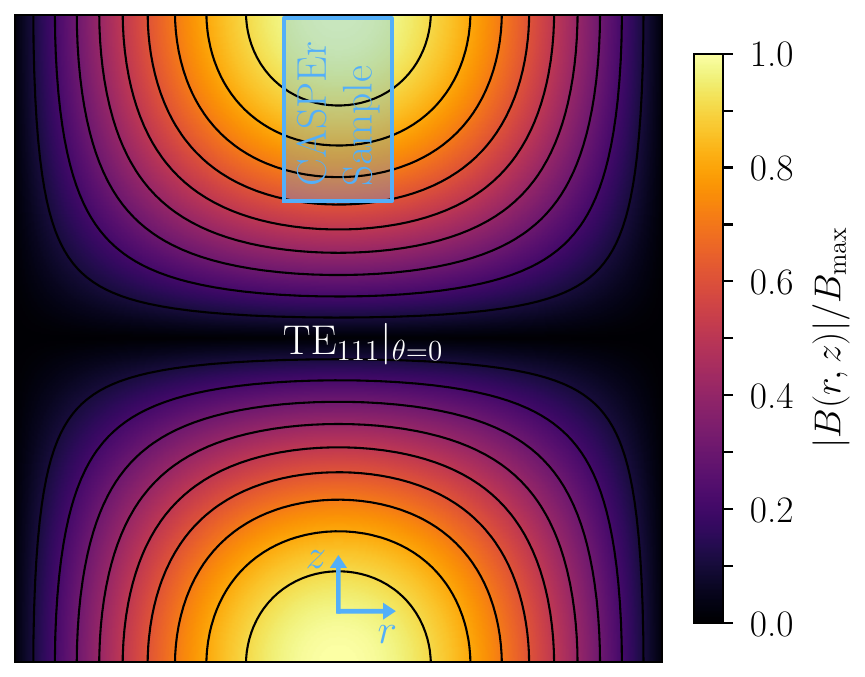}\hspace{0.1cm}
\includegraphics[height=0.273\textwidth]{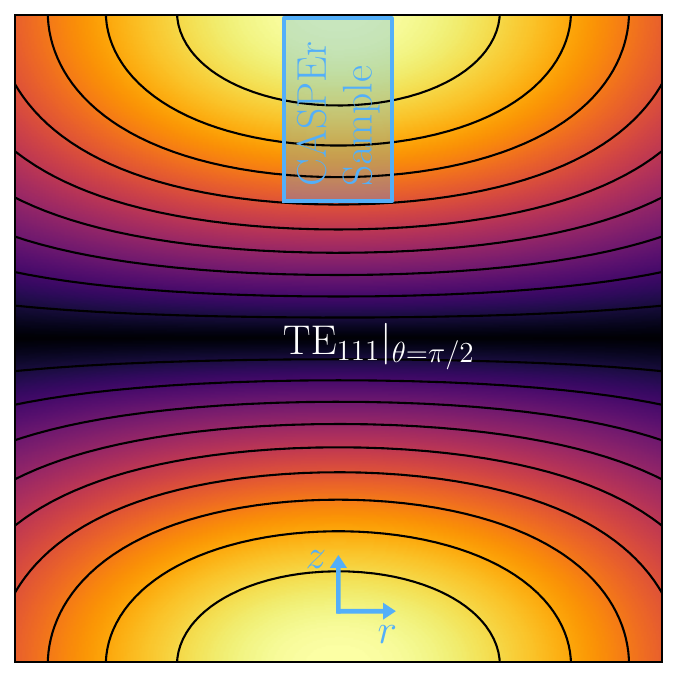}
\vspace{-0.3cm}
\caption{The profile of the physical magnetic fields DM could excite as compared the size of CASPEr's sample of nuclear spins.
The contours denote 10\% changes in the magnetic field value.
The dominant mode the axion-photon coupling can excite is TM$_{010}$, whereas a dark photon can also excite the TE$_{111}$ mode depending upon its polarization.
The figure highlights that as the sample is moved within the magnetic field the response varies, opening a path to distinguishing different DM signals and even mapping out the dark photon polarization.
}
\vspace{-0.5cm}
\label{fig:Placement}
\end{figure*}

This effective dark-photon current penetrates shielded regions, such as surrounds CASPEr-Gradient.
Taking the shielded region to have parametric size $L$, DM NMR operates in the regime where $mL \ll 1$ and therefore the induced magnetic field dominates~\cite{Chaudhuri:2014dla}.
In that limit we can neglect the displacement current implying the relevant equation is $\nabla \times \bH \simeq - \epsilon m^2 \bA'$, confirming the parametric size of the induced magnetic field as $\epsilon m^2 L \bA'$.
From Eq.~\eqref{eq:HDM-sens}, the parametric sensitivity scales as
\be
\epsilon \sim \frac{v}{\gamma L} \sqrt{\frac{1}{m \rho_{\DM} n T_2 V}}.
\label{eq:DP-sens}
\ee
When compared to the scaling for $g_{\sN}$ this result is consistent with Eq.~\eqref{eq:eps-qual}.

To fully quantify the sensitivity we must consider how the modes of the shielded region can be driven by the dark-photon current and the geometry of the associated magnetic fields.
We treat the shielded region as a cylindrical cavity (deviations away from this will not strongly impact our projections) and as $mL \ll 1$ the DM driving frequency is always much smaller than the electronic resonant frequencies ($\w_0 \sim 1/L$), implying we can ignore dissipative effects.
Further, the precise composition of the shielding material is not important, as the DM signal is at sufficiently high frequencies that the shield can be treated as a perfect conductor.
We can then decompose the cavity field into the conventional transverse electric (TE) and transverse magnetic (TM) modes, each labelled by indices $\ell = (m,n,p)$ (for the azimuthal, radial, and longitudinal indices), from which the amount each mode is excited by the dark photon can be computed directly.
The computation is standard and therefore not presented here, although the full details are provided in the SM.

One aspect of the calculation that must be mentioned, however, is the spatial profile of the dark-photon induced magnetic field.
This dictates the direction of the magnetic field experienced by the spin sample and opens the possibility of distinguishing various DM induced NMR signals; see Fig.~\ref{fig:Placement}.
To understand the field profile, we first emphasise that it is the lowest accessible cavity modes that are dominantly excited.
If $\bE_{\ell}$ represents the field of a given mode, then the induced magnetic field depends on the following integral over the shielded volume,
\be
\int dV\, \bE_{\ell}^* \cdot \bJ_{\rm eff} \simeq - \epsilon m^2 \bA' \cdot \int dV\, \bE_{\ell}^*,
\label{eq:Jeff-coup}
\ee
where the accuracy of the approximation exploits the spatial uniformity of DM over spatial scales ${\cal O}(L)$~\cite{Cheong:2024ose}.
The value of the final integral is suppressed for higher modes; indeed, it is an excellent approximation to only retain the lowest non-vanishing modes, TM$_{010}$ driven by $A'_z$ and TE$_{111}$ driven by $A'_{x,y}$.
Accordingly, the modes driven are sensitive to the polarization of the dark photon, a fact that could be exploited to not only measure the polarization distribution in the event of a detection but also to distinguish the signal from axion DM.

The spatial profile of the modes determines the ideal placement of the spin sample as shown in Fig.~\ref{fig:Placement}.
For CASPEr-Gradient the sample is sufficiently smaller than the shielded region such that the sample can in principle be moved around and in the case of the TM$_{010}$ mode can inhabit a region of uniform induced field.
(For other modes and a larger sample the effect of a non-uniform field across the sample must be accounted for, we consider this scenario in the SM.)
In detail, we take the shielding to have a radius of $R \simeq9$\,cm and height of 18\,cm, whereas the cylindrical sample has a height of 3\,cm and radius of 2.5\,cm.
In Fig.~\ref{fig:Placement} the axis of the sample is taken to be perpendicular to that of the shielding.
Generally, the optimal placement is determined by looking for where the modes take their maximum value; for the TM$_{010}$ mode the sample should be placed at a radius of roughly $0.77 R$, whereas for the TE$_{111}$ mode the sample should be placed on-axis radially and near the end caps longitudinally.

Comparing Eq.~\eqref{eq:DP-sens} to Fig.~\ref{fig:Results} we find that the parametric estimate for fixed $T_2$ accurately reflects the scaling of the full result, but overstates the sensitivity by roughly a decade.
The difference is a combination of ${\cal O}(1)$ factors and roughly a factor of $\sim$$1.8$ suppression from the geometry of the cavity.
The sensitivity breaks down for $m \gtrsim 1\,\mu$eV as the cavity modes begin to be driven resonantly.
This breakdown is largely irrelevant to our projections because the largest mass value that can be observed is dictated by the externally applied magnetic field and is below the point where the $Q$-factor of the cavity must be specified.
The weaker projections arise by assuming that the scan is limited by a part per million line-width at every frequency due to magnetic field inhomogeneities~\cite{Walter:2023wto}.
The effective $T_2$ is then frequency dependent, $T_2 \propto 1/m$, so that Eq.~\eqref{eq:DP-sens} predicts a flat sensitivity scaling with mass as approximately observed.

\vspace{0.2cm}
\noindent {\bf Axion-Photon NMR.}
%
The considerations for axion DM that couples through ${\cal L}_{a\gamma\gamma}$ are almost identical to those for the dark-photon.
For axion DM, the leading effect is an induced current $\bJ_{\rm eff} \simeq \gagg (\pd_t a) \bB$.
As the background magnetic field points along the axis of the cylindrical shielded region, from Eq.~\eqref{eq:Jeff-coup} it is apparent that the axion does not couple to the TE modes, leaving the dominant contribution from the TM$_{010}$ mode, again as shown in Fig.~\ref{fig:Placement}.
For Fig.~\ref{fig:Results} we assume the sample is placed optimally.
The quantitative sensitivity there can be determined as for the dark photon: $H_{\DM} \simeq \gagg B_0 L (\pd_t a)$.
A critical point, however, is that $B_0$ is not a fixed parameter, but is rather adjusted to scan through various DM masses.
If we denote $\bar{B}_0$ as the magnetic field used to scan the largest mass in the search, $\bar{m}$, then $B_0 = \bar{B}_0 (m/\bar{m})$.
Finally, we can use Eq.~\eqref{eq:HDM-sens} to obtain a parametric sensitivity of
\be
\gagg \sim \frac{v \bar{m}}{\gamma \bar{B}_0 L} \sqrt{\frac{1}{m\rho_{\DM} nT_2V}}.
\ee
As for the dark photon, this result correctly reproduces the scaling in Fig.~\ref{fig:Results}.

\vspace{0.2cm}
\noindent {\bf Discussion.}
%
Axion NMR is a promising avenue to search for axion DM through its coupling to nucleons.
This coupling generates an \emph{effective} magnetic field that couples to a spin sample, causing it to precess in a measurable manner.
Significant experimental efforts are underway by the CASPEr-Gradient collaboration~\cite{Graham:2013gfa, Budker:2013hfa, Garcon:2017ixh, JacksonKimball:2017elr}, and other proposals have been made to conduct similar experiments~\cite{Gao:2022nuq,Foster:2023bxl,Brandenstein:2022eif,Bloch:2019lcy}.
Axion NMR is also an active area of theoretical research aiming to refine the ultimate limits in sensitivity achievable with this technique~\cite{Dror:2022xpi,Aybas:2021cdk,Zhang_2023}.

In this Letter we quantified the sensitivity that axion NMR experiments have to alternative DM signals that produce \emph{real} magnetic fields in the vicinity of the hyper-polarised spin sample.
We demonstrated that CASPEr-Gradient can have world-leading sensitivity to dark-photon DM as well as to the axion-photon coupling in the same experimental setup they use to search for the axion-nucleon coupling.
Importantly, the three signals can be distinguished by virtue of their symmetry properties.
The canonical signal induced by the axion-nucleon coupling is homogeneous throughout the electromagnetically-shielded volume in which the spin sample is placed.
In contrast, the dark photon and axion-photon coupling generate inhomogeneous signals: Fig.~\ref{fig:Placement} shows that the three signals can be differentiated by modifying the placement of the sample inside the shield.

The coming years promise revolutionary progress in the search for ultralight DM with masses below a $\mu$eV.
In particular, as instruments like DMRadio~\cite{DMRadio:2022jfv,DMRadio:2022pkf} or heterodyne SRF haloscopes~\cite{Berlin:2019ahk,Berlin:2020vrk,Berlin:2022hfx} are increased in scale, the goal is to improve the sensitivity by many, many orders of magnitude beyond existing low-mass haloscopes~\cite{Ouellet:2018beu,Ouellet:2019tlz,Salemi:2021gck,Gramolin:2020ict} to push towards the QCD axion prediction.
DM NMR represents a completely distinct experimental strategy to push towards that same goal, and as the present work has highlighted it can probe many forms of DM as it does so.
For this reason it will be critical to determine how far the DM NMR approach can ultimately be scaled, in particular understanding the quantum mechanics-imposed limitations~\cite{Aybas:2021cdk} and ways they can be evaded through the use of quantum resources~\cite{Boyers:2025qgc}.

\medskip
\noindent {\it Acknowledgments.}
%
We thank the CASPEr collaboration for both feedback and providing technical regarding the CASPEr-Gradient instrument.
The work of CB and SARE was supported by Swiss National Science Foundation (SNSF) Ambizione grant PZ00P2 193322, \textit{New frontiers from sub-eV to super-TeV}.
JML was co-funded by the European Union and supported by the Czech Ministry of Education, Youth and Sports (Project No. FORTE–CZ.02.01.01/00/22 008/0004632), as well as by the Deutsche Forschungsgemeinschaft under Germany’s Excellence Strategy - EXC 2121 “Quantum Universe” - 390833306.
The research of NLR was supported by the Office of High Energy Physics of the U.S. Department of Energy under contract DE-AC02-05CH11231.

\bibliographystyle{utphys}
\bibliography{Bibliography}

\providecommand{\href}[2]{#2}\begingroup\raggedright\begin{thebibliography}{10}

\bibitem{Cheong:2024ose}
D.~Y. Cheong, N.~L. Rodd, and L.-T. Wang, ``{Quantum description of wave dark
  matter},'' \href{https://dx.doi.org/10.1103/PhysRevD.111.015028}{{\em Phys.
  Rev. D} {\bfseries 111} no.~1, (2025) 015028},
  \href{https://arxiv.org/abs/2408.04696}{{\ttfamily arXiv:2408.04696
  [hep-ph]}}.

\bibitem{Dror:2021nyr}
J.~A. Dror, H.~Murayama, and N.~L. Rodd, ``{Cosmic axion background},''
  \href{https://dx.doi.org/10.1103/PhysRevD.103.115004}{{\em Phys. Rev. D}
  {\bfseries 103} no.~11, (2021) 115004},
  \href{https://arxiv.org/abs/2101.09287}{{\ttfamily arXiv:2101.09287
  [hep-ph]}}. [Erratum: Phys.Rev.D 106, 119902 (2022)].

\bibitem{ADMX:2023rsk}
{\bfseries ADMX} Collaboration, T.~Nitta {\em et~al.}, ``{Search for a
  Dark-Matter-Induced Cosmic Axion Background with ADMX},''
  \href{https://dx.doi.org/10.1103/PhysRevLett.131.101002}{{\em Phys. Rev.
  Lett.} {\bfseries 131} no.~10, (2023) 101002},
  \href{https://arxiv.org/abs/2303.06282}{{\ttfamily arXiv:2303.06282
  [hep-ex]}}.

\bibitem{Berlin:2021txa}
A.~Berlin, D.~Blas, R.~Tito~D'Agnolo, S.~A.~R. Ellis, R.~Harnik, Y.~Kahn, and
  J.~Sch\"utte-Engel, ``{Detecting high-frequency gravitational waves with
  microwave cavities},''
  \href{https://dx.doi.org/10.1103/PhysRevD.105.116011}{{\em Phys. Rev. D}
  {\bfseries 105} no.~11, (2022) 116011},
  \href{https://arxiv.org/abs/2112.11465}{{\ttfamily arXiv:2112.11465
  [hep-ph]}}.

\bibitem{Berlin:2023grv}
A.~Berlin, D.~Blas, R.~Tito~D'Agnolo, S.~A.~R. Ellis, R.~Harnik, Y.~Kahn,
  J.~Sch\"utte-Engel, and M.~Wentzel, ``{Electromagnetic cavities as mechanical
  bars for gravitational waves},''
  \href{https://dx.doi.org/10.1103/PhysRevD.108.084058}{{\em Phys. Rev. D}
  {\bfseries 108} no.~8, (2023) 084058},
  \href{https://arxiv.org/abs/2303.01518}{{\ttfamily arXiv:2303.01518
  [hep-ph]}}.

\bibitem{Domcke:2022rgu}
V.~Domcke, C.~Garcia-Cely, and N.~L. Rodd, ``{Novel Search for High-Frequency
  Gravitational Waves with Low-Mass Axion Haloscopes},''
  \href{https://dx.doi.org/10.1103/PhysRevLett.129.041101}{{\em Phys. Rev.
  Lett.} {\bfseries 129} no.~4, (2022) 041101},
  \href{https://arxiv.org/abs/2202.00695}{{\ttfamily arXiv:2202.00695
  [hep-ph]}}.

\bibitem{Domcke:2023bat}
V.~Domcke, C.~Garcia-Cely, S.~M. Lee, and N.~L. Rodd, ``{Symmetries and
  selection rules: optimising axion haloscopes for Gravitational Wave
  searches},'' \href{https://dx.doi.org/10.1007/JHEP03(2024)128}{{\em JHEP}
  {\bfseries 03} (2024) 128},
  \href{https://arxiv.org/abs/2306.03125}{{\ttfamily arXiv:2306.03125
  [hep-ph]}}.

\bibitem{Domcke:2024mfu}
V.~Domcke, S.~A.~R. Ellis, and N.~L. Rodd, ``{Magnets are Weber Bar
  Gravitational Wave Detectors},''
  \href{https://arxiv.org/abs/2408.01483}{{\ttfamily arXiv:2408.01483
  [hep-ph]}}.

\bibitem{Pappas:2025zld}
K.~M.~W. Pappas {\em et~al.}, ``{High-Frequency Gravitational Wave Search with
  ABRACADABRA-10 cm},'' \href{https://arxiv.org/abs/2505.02821}{{\ttfamily
  arXiv:2505.02821 [hep-ex]}}.

\bibitem{TitoDAgnolo:2024uku}
R.~Tito~D'Agnolo and S.~A.~R. Ellis, ``{Classical (and quantum) heuristics for
  gravitational wave detection},''
  \href{https://dx.doi.org/10.1007/JHEP04(2025)164}{{\em JHEP} {\bfseries 04}
  (2025) 164}, \href{https://arxiv.org/abs/2412.17897}{{\ttfamily
  arXiv:2412.17897 [gr-qc]}}.

\bibitem{Graham:2013gfa}
P.~W. Graham and S.~Rajendran, ``{New Observables for Direct Detection of Axion
  Dark Matter},'' \href{https://dx.doi.org/10.1103/PhysRevD.88.035023}{{\em
  Phys. Rev. D} {\bfseries 88} (2013) 035023},
  \href{https://arxiv.org/abs/1306.6088}{{\ttfamily arXiv:1306.6088 [hep-ph]}}.

\bibitem{Budker:2013hfa}
D.~Budker, P.~W. Graham, M.~Ledbetter, S.~Rajendran, and A.~Sushkov,
  ``{Proposal for a Cosmic Axion Spin Precession Experiment (CASPEr)},''
  \href{https://dx.doi.org/10.1103/PhysRevX.4.021030}{{\em Phys. Rev. X}
  {\bfseries 4} no.~2, (2014) 021030},
  \href{https://arxiv.org/abs/1306.6089}{{\ttfamily arXiv:1306.6089 [hep-ph]}}.

\bibitem{Kalia:2024eml}
S.~Kalia, D.~Budker, D.~F.~J. Kimball, W.~Ji, Z.~Liu, A.~O. Sushkov,
  C.~Timberlake, H.~Ulbricht, A.~Vinante, and T.~Wang, ``{Ultralight dark
  matter detection with levitated ferromagnets},''
  \href{https://dx.doi.org/10.1103/PhysRevD.110.115029}{{\em Phys. Rev. D}
  {\bfseries 110} no.~11, (2024) 115029},
  \href{https://arxiv.org/abs/2408.15330}{{\ttfamily arXiv:2408.15330
  [hep-ph]}}.

\bibitem{Dror:2022xpi}
J.~A. Dror, S.~Gori, J.~M. Leedom, and N.~L. Rodd, ``{Sensitivity of
  Spin-Precession Axion Experiments},''
  \href{https://dx.doi.org/10.1103/PhysRevLett.130.181801}{{\em Phys. Rev.
  Lett.} {\bfseries 130} no.~18, (2023) 181801},
  \href{https://arxiv.org/abs/2210.06481}{{\ttfamily arXiv:2210.06481
  [hep-ph]}}.

\bibitem{Holdom:1985ag}
B.~Holdom, ``{Two U(1)'s and Epsilon Charge Shifts},''
  \href{https://dx.doi.org/10.1016/0370-2693(86)91377-8}{{\em Phys. Lett. B}
  {\bfseries 166} (1986) 196--198}.

\bibitem{Chaudhuri:2014dla}
S.~Chaudhuri, P.~W. Graham, K.~Irwin, J.~Mardon, S.~Rajendran, and Y.~Zhao,
  ``{Radio for hidden-photon dark matter detection},''
  \href{https://dx.doi.org/10.1103/PhysRevD.92.075012}{{\em Phys. Rev. D}
  {\bfseries 92} no.~7, (2015) 075012},
  \href{https://arxiv.org/abs/1411.7382}{{\ttfamily arXiv:1411.7382 [hep-ph]}}.

\bibitem{Caputo:2021eaa}
A.~Caputo, A.~J. Millar, C.~A.~J. O'Hare, and E.~Vitagliano, ``{Dark photon
  limits: A handbook},''
  \href{https://dx.doi.org/10.1103/PhysRevD.104.095029}{{\em Phys. Rev. D}
  {\bfseries 104} no.~9, (2021) 095029},
  \href{https://arxiv.org/abs/2105.04565}{{\ttfamily arXiv:2105.04565
  [hep-ph]}}.

\bibitem{JacksonKimball:2017elr}
D.~F. Jackson~Kimball {\em et~al.}, ``{Overview of the Cosmic Axion Spin
  Precession Experiment (CASPEr)},''
  \href{https://dx.doi.org/10.1007/978-3-030-43761-9_13}{{\em Springer Proc.
  Phys.} {\bfseries 245} (2020) 105--121},
  \href{https://arxiv.org/abs/1711.08999}{{\ttfamily arXiv:1711.08999
  [physics.ins-det]}}.

\bibitem{An:2024wmc}
H.~An, S.~Ge, J.~Liu, and M.~Liu, ``{In Situ Measurements of Dark Photon Dark
  Matter Using Parker Solar Probe: Going beyond the Radio Window},''
  \href{https://dx.doi.org/10.1103/PhysRevLett.134.171001}{{\em Phys. Rev.
  Lett.} {\bfseries 134} no.~17, (2025) 171001},
  \href{https://arxiv.org/abs/2405.12285}{{\ttfamily arXiv:2405.12285
  [hep-ph]}}.

\bibitem{An:2023wij}
H.~An, X.~Chen, S.~Ge, J.~Liu, and Y.~Luo, ``{Searching for ultralight dark
  matter conversion in solar corona using Low Frequency Array data},''
  \href{https://dx.doi.org/10.1038/s41467-024-45033-4}{{\em Nature Commun.}
  {\bfseries 15} no.~1, (2024) 915},
  \href{https://arxiv.org/abs/2301.03622}{{\ttfamily arXiv:2301.03622
  [hep-ph]}}.

\bibitem{Paola_Arias_2012}
P.~Arias, D.~Cadamuro, M.~Goodsell, J.~Jaeckel, J.~Redondo, and A.~Ringwald,
  ``{WISPy Cold Dark Matter},''
  \href{https://dx.doi.org/10.1088/1475-7516/2012/06/013}{{\em JCAP} {\bfseries
  06} (2012) 013}, \href{https://arxiv.org/abs/1201.5902}{{\ttfamily
  arXiv:1201.5902 [hep-ph]}}.

\bibitem{Caputo_2020}
A.~Caputo, H.~Liu, S.~Mishra-Sharma, and J.~T. Ruderman, ``{Dark Photon
  Oscillations in Our Inhomogeneous Universe},''
  \href{https://dx.doi.org/10.1103/PhysRevLett.125.221303}{{\em Phys. Rev.
  Lett.} {\bfseries 125} no.~22, (2020) 221303},
  \href{https://arxiv.org/abs/2002.05165}{{\ttfamily arXiv:2002.05165
  [astro-ph.CO]}}.

\bibitem{McDermott_2020}
S.~D. McDermott and S.~J. Witte, ``{Cosmological evolution of light dark photon
  dark matter},'' \href{https://dx.doi.org/10.1103/PhysRevD.101.063030}{{\em
  Phys. Rev. D} {\bfseries 101} no.~6, (2020) 063030},
  \href{https://arxiv.org/abs/1911.05086}{{\ttfamily arXiv:1911.05086
  [hep-ph]}}.

\bibitem{PhysRevLett.105.171801}
{\bfseries ADMX} Collaboration, A.~Wagner {\em et~al.}, ``{A Search for Hidden
  Sector Photons with ADMX},''
  \href{https://dx.doi.org/10.1103/PhysRevLett.105.171801}{{\em Phys. Rev.
  Lett.} {\bfseries 105} (2010) 171801},
  \href{https://arxiv.org/abs/1007.3766}{{\ttfamily arXiv:1007.3766 [hep-ex]}}.

\bibitem{Nguyen_2019}
L.~H. Nguyen, A.~Lobanov, and D.~Horns, ``{First results from the WISPDMX radio
  frequency cavity searches for hidden photon dark matter},''
  \href{https://dx.doi.org/10.1088/1475-7516/2019/10/014}{{\em JCAP} {\bfseries
  10} (2019) 014}, \href{https://arxiv.org/abs/1907.12449}{{\ttfamily
  arXiv:1907.12449 [hep-ex]}}.

\bibitem{PhysRevLett.124.101802}
S.~Lee, S.~Ahn, J.~Choi, B.~R. Ko, and Y.~K. Semertzidis, ``{Axion Dark Matter
  Search around 6.7 $\mu$eV},''
  \href{https://dx.doi.org/10.1103/PhysRevLett.124.101802}{{\em Phys. Rev.
  Lett.} {\bfseries 124} no.~10, (2020) 101802},
  \href{https://arxiv.org/abs/2001.05102}{{\ttfamily arXiv:2001.05102
  [hep-ex]}}.

\bibitem{Zhong_2018}
{\bfseries HAYSTAC} Collaboration, L.~Zhong {\em et~al.}, ``{Results from phase
  1 of the HAYSTAC microwave cavity axion experiment},''
  \href{https://dx.doi.org/10.1103/PhysRevD.97.092001}{{\em Phys. Rev. D}
  {\bfseries 97} no.~9, (2018) 092001},
  \href{https://arxiv.org/abs/1803.03690}{{\ttfamily arXiv:1803.03690
  [hep-ex]}}.

\bibitem{Backes_2021}
{\bfseries HAYSTAC} Collaboration, K.~M. Backes {\em et~al.}, ``{A
  quantum-enhanced search for dark matter axions},''
  \href{https://dx.doi.org/10.1038/s41586-021-03226-7}{{\em Nature} {\bfseries
  590} no.~7845, (2021) 238--242},
  \href{https://arxiv.org/abs/2008.01853}{{\ttfamily arXiv:2008.01853
  [quant-ph]}}.

\bibitem{PhysRevD.104.012013}
B.~Godfrey {\em et~al.}, ``{Search for dark photon dark matter: Dark E field
  radio pilot experiment},''
  \href{https://dx.doi.org/10.1103/PhysRevD.104.012013}{{\em Phys. Rev. D}
  {\bfseries 104} no.~1, (2021) 012013},
  \href{https://arxiv.org/abs/2101.02805}{{\ttfamily arXiv:2101.02805
  [physics.ins-det]}}.

\bibitem{ahn2024extensivesearchaxiondark}
{\bfseries CAPP} Collaboration, S.~Ahn {\em et~al.}, ``{Extensive Search for
  Axion Dark Matter over 1~GHz with CAPP\textquoteright{}S Main Axion
  Experiment},'' \href{https://dx.doi.org/10.1103/PhysRevX.14.031023}{{\em
  Phys. Rev. X} {\bfseries 14} no.~3, (2024) 031023},
  \href{https://arxiv.org/abs/2402.12892}{{\ttfamily arXiv:2402.12892
  [hep-ex]}}.

\bibitem{Alesini:2020vny}
D.~Alesini {\em et~al.}, ``{Search for invisible axion dark matter of mass
  m$_a=43~\mu$eV with the QUAX--$a\gamma$ experiment},''
  \href{https://dx.doi.org/10.1103/PhysRevD.103.102004}{{\em Phys. Rev. D}
  {\bfseries 103} no.~10, (2021) 102004},
  \href{https://arxiv.org/abs/2012.09498}{{\ttfamily arXiv:2012.09498
  [hep-ex]}}.

\bibitem{Ahyoune:2024klt}
S.~Ahyoune {\em et~al.}, ``{RADES axion search results with a high-temperature
  superconducting cavity in an 11.7 T magnet},''
  \href{https://dx.doi.org/10.1007/JHEP04(2025)113}{{\em JHEP} {\bfseries 04}
  (2025) 113}, \href{https://arxiv.org/abs/2403.07790}{{\ttfamily
  arXiv:2403.07790 [hep-ex]}}.

\bibitem{Quiskamp_2022}
A.~P. Quiskamp, B.~T. McAllister, P.~Altin, E.~N. Ivanov, M.~Goryachev, and
  M.~E. Tobar, ``{Direct search for dark matter axions excluding ALP cogenesis
  in the 63- to 67-\ensuremath{\mu}eV range with the ORGAN experiment},''
  \href{https://dx.doi.org/10.1126/sciadv.abq3765}{{\em Sci. Adv.} {\bfseries
  8} no.~27, (2022) abq3765},
  \href{https://arxiv.org/abs/2203.12152}{{\ttfamily arXiv:2203.12152
  [hep-ex]}}.

\bibitem{Anastassopoulos2017}
{\bfseries CAST} Collaboration, V.~Anastassopoulos {\em et~al.}, ``{New CAST
  Limit on the Axion-Photon Interaction},''
  \href{https://dx.doi.org/10.1038/nphys4109}{{\em Nature Phys.} {\bfseries 13}
  (2017) 584--590}, \href{https://arxiv.org/abs/1705.02290}{{\ttfamily
  arXiv:1705.02290 [hep-ex]}}.

\bibitem{PhysRevLett.131.111004}
D.~Noordhuis, A.~Prabhu, S.~J. Witte, A.~Y. Chen, F.~Cruz, and C.~Weniger,
  ``{Novel Constraints on Axions Produced in Pulsar Polar-Cap Cascades},''
  \href{https://dx.doi.org/10.1103/PhysRevLett.131.111004}{{\em Phys. Rev.
  Lett.} {\bfseries 131} no.~11, (2023) 111004},
  \href{https://arxiv.org/abs/2209.09917}{{\ttfamily arXiv:2209.09917
  [hep-ph]}}.

\bibitem{PhysRevD.105.103034}
C.~Dessert, D.~Dunsky, and B.~R. Safdi, ``{Upper limit on the axion-photon
  coupling from magnetic white dwarf polarization},''
  \href{https://dx.doi.org/10.1103/PhysRevD.105.103034}{{\em Phys. Rev. D}
  {\bfseries 105} no.~10, (2022) 103034},
  \href{https://arxiv.org/abs/2203.04319}{{\ttfamily arXiv:2203.04319
  [hep-ph]}}.

\bibitem{PhysRevD.107.083027}
J.~Davies, M.~Meyer, and G.~Cotter, ``{Constraints on axionlike particles from
  a combined analysis of three flaring Fermi flat-spectrum radio quasars},''
  \href{https://dx.doi.org/10.1103/PhysRevD.107.083027}{{\em Phys. Rev. D}
  {\bfseries 107} no.~8, (2023) 083027},
  \href{https://arxiv.org/abs/2211.03414}{{\ttfamily arXiv:2211.03414
  [astro-ph.HE]}}.

\bibitem{ABE2024101425}
{\bfseries MAGIC} Collaboration, H.~Abe {\em et~al.}, ``{Constraints on
  axion-like particles with the Perseus Galaxy Cluster with MAGIC},''
  \href{https://dx.doi.org/10.1016/j.dark.2024.101425}{{\em Phys. Dark Univ.}
  {\bfseries 44} (2024) 101425},
  \href{https://arxiv.org/abs/2401.07798}{{\ttfamily arXiv:2401.07798
  [astro-ph.HE]}}.

\bibitem{PhysRevLett.116.161101}
{\bfseries Fermi-LAT} Collaboration, M.~Ajello {\em et~al.}, ``{Search for
  Spectral Irregularities due to Photon\textendash{}Axionlike-Particle
  Oscillations with the Fermi Large Area Telescope},''
  \href{https://dx.doi.org/10.1103/PhysRevLett.116.161101}{{\em Phys. Rev.
  Lett.} {\bfseries 116} no.~16, (2016) 161101},
  \href{https://arxiv.org/abs/1603.06978}{{\ttfamily arXiv:1603.06978
  [astro-ph.HE]}}.

\bibitem{Benabou:2025jcv}
J.~N. Benabou, C.~Dessert, K.~C. Patra, T.~G. Brink, W.~Zheng, A.~V.
  Filippenko, and B.~R. Safdi, ``{Search for Axions in Magnetic White Dwarf
  Polarization at Lick and Keck Observatories},''
  \href{https://arxiv.org/abs/2504.12377}{{\ttfamily arXiv:2504.12377
  [hep-ph]}}.

\bibitem{Dine:1981rt}
M.~Dine, W.~Fischler, and M.~Srednicki, ``{A Simple Solution to the Strong CP
  Problem with a Harmless Axion},''
  \href{https://dx.doi.org/10.1016/0370-2693(81)90590-6}{{\em Phys. Lett. B}
  {\bfseries 104} (1981) 199--202}.

\bibitem{PhysRevLett.43.103}
J.~E. Kim, ``{Weak Interaction Singlet and Strong CP Invariance},''
  \href{https://dx.doi.org/10.1103/PhysRevLett.43.103}{{\em Phys. Rev. Lett.}
  {\bfseries 43} (1979) 103}.

\bibitem{Shifman:1979if}
M.~A. Shifman, A.~I. Vainshtein, and V.~I. Zakharov, ``{Can Confinement Ensure
  Natural CP Invariance of Strong Interactions?},''
  \href{https://dx.doi.org/10.1016/0550-3213(80)90209-6}{{\em Nucl. Phys. B}
  {\bfseries 166} (1980) 493--506}.

\bibitem{Co_2021}
R.~T. Co, L.~J. Hall, and K.~Harigaya, ``{Predictions for Axion Couplings from
  ALP Cogenesis},'' \href{https://dx.doi.org/10.1007/JHEP01(2021)172}{{\em
  JHEP} {\bfseries 01} (2021) 172},
  \href{https://arxiv.org/abs/2006.04809}{{\ttfamily arXiv:2006.04809
  [hep-ph]}}.

\bibitem{AxionLimits}
C.~O'Hare, ``cajohare/axionlimits: Axionlimits,''
  \url{https://cajohare.github.io/AxionLimits/}, July, 2020.

\bibitem{Cyncynates:2024yxm}
D.~Cyncynates and Z.~J. Weiner, ``{Experimental targets for dark photon dark
  matter},'' \href{https://arxiv.org/abs/2410.14774}{{\ttfamily
  arXiv:2410.14774 [hep-ph]}}.

\bibitem{East:2022rsi}
W.~E. East and J.~Huang, ``{Dark photon vortex formation and dynamics},''
  \href{https://dx.doi.org/10.1007/JHEP12(2022)089}{{\em JHEP} {\bfseries 12}
  (2022) 089}, \href{https://arxiv.org/abs/2206.12432}{{\ttfamily
  arXiv:2206.12432 [hep-ph]}}.

\bibitem{Walter:2023wto}
J.~Walter, H.~Bekker, J.~Blanchard, D.~Budker, N.~L. Figueroa, A.~Wickenbrock,
  Y.~Zhang, and P.~Zhou, ``{Fast shimming algorithm based on Bayesian
  optimization for magnetic resonance based dark matter search},''
  \href{https://arxiv.org/abs/2309.11614}{{\ttfamily arXiv:2309.11614
  [astro-ph.CO]}}.

\bibitem{Garcon:2017ixh}
A.~Garcon {\em et~al.}, ``{The cosmic axion spin precession experiment
  (CASPEr): a dark-matter search with nuclear magnetic resonance},''
  \href{https://dx.doi.org/10.1088/2058-9565/aa9861}{{\em Quantum Sci.
  Technol.} {\bfseries 3} no.~1, (2017) 014008},
  \href{https://arxiv.org/abs/1707.05312}{{\ttfamily arXiv:1707.05312
  [physics.ins-det]}}.

\bibitem{Gao:2022nuq}
C.~Gao, W.~Halperin, Y.~Kahn, M.~Nguyen, J.~Sch\"utte-Engel, and J.~W. Scott,
  ``{Axion Wind Detection with the Homogeneous Precession Domain of Superfluid
  Helium-3},'' \href{https://dx.doi.org/10.1103/PhysRevLett.129.211801}{{\em
  Phys. Rev. Lett.} {\bfseries 129} no.~21, (2022) 211801},
  \href{https://arxiv.org/abs/2208.14454}{{\ttfamily arXiv:2208.14454
  [hep-ph]}}.

\bibitem{Foster:2023bxl}
J.~W. Foster, C.~Gao, W.~Halperin, Y.~Kahn, A.~Mande, M.~Nguyen,
  J.~Sch\"utte-Engel, and J.~W. Scott, ``{Statistics and sensitivity of axion
  wind detection with the homogeneous precession domain of superfluid
  helium-3},'' \href{https://dx.doi.org/10.1103/PhysRevD.110.115020}{{\em Phys.
  Rev. D} {\bfseries 110} no.~11, (2024) 115020},
  \href{https://arxiv.org/abs/2310.07791}{{\ttfamily arXiv:2310.07791
  [hep-ph]}}.

\bibitem{Brandenstein:2022eif}
C.~Brandenstein, S.~Stelzl, E.~Gutsmiedl, W.~Schott, A.~Weiler, and
  P.~Fierlinger, ``{Towards an electrostatic storage ring for fundamental
  physics measurements},''
  \href{https://dx.doi.org/10.1051/epjconf/202328201017}{{\em EPJ Web Conf.}
  {\bfseries 282} (2023) 01017},
  \href{https://arxiv.org/abs/2211.08439}{{\ttfamily arXiv:2211.08439
  [hep-ex]}}.

\bibitem{Bloch:2019lcy}
I.~M. Bloch, Y.~Hochberg, E.~Kuflik, and T.~Volansky, ``{Axion-like Relics: New
  Constraints from Old Comagnetometer Data},''
  \href{https://dx.doi.org/10.1007/JHEP01(2020)167}{{\em JHEP} {\bfseries 01}
  (2020) 167}, \href{https://arxiv.org/abs/1907.03767}{{\ttfamily
  arXiv:1907.03767 [hep-ph]}}.

\bibitem{Aybas:2021cdk}
D.~Aybas, H.~Bekker, J.~W. Blanchard, D.~Budker, G.~P. Centers, N.~L. Figueroa,
  A.~V. Gramolin, D.~F.~J. Kimball, A.~Wickenbrock, and A.~O. Sushkov,
  ``{Quantum sensitivity limits of nuclear magnetic resonance experiments
  searching for new fundamental physics},''
  \href{https://dx.doi.org/10.1088/2058-9565/abfbbc}{{\em Quantum Sci.
  Technol.} {\bfseries 6} no.~3, (2021) 034007},
  \href{https://arxiv.org/abs/2103.06284}{{\ttfamily arXiv:2103.06284
  [quant-ph]}}.

\bibitem{Zhang_2023}
Y.~Zhang, D.~A. Tumturk, H.~Bekker, D.~Budker, D.~F.~J. Kimball, A.~O. Sushkov,
  and A.~Wickenbrock, ``Frequency‐scanning considerations in axionlike dark
  matter spin‐precession experiments,''
  \href{https://dx.doi.org/10.1002/andp.202300223}{{\em Annalen der Physik}
  {\bfseries 536} no.~1, (2023) 2300223}.

\bibitem{DMRadio:2022jfv}
{\bfseries DMRadio} Collaboration, L.~Brouwer {\em et~al.}, ``{Proposal for a
  definitive search for GUT-scale QCD axions},''
  \href{https://dx.doi.org/10.1103/PhysRevD.106.112003}{{\em Phys. Rev. D}
  {\bfseries 106} no.~11, (2022) 112003},
  \href{https://arxiv.org/abs/2203.11246}{{\ttfamily arXiv:2203.11246
  [hep-ex]}}.

\bibitem{DMRadio:2022pkf}
{\bfseries DMRadio} Collaboration, L.~Brouwer {\em et~al.}, ``{Projected
  sensitivity of DMRadio-m3: A search for the QCD axion below
  1\,\,\ensuremath{\mu}eV},''
  \href{https://dx.doi.org/10.1103/PhysRevD.106.103008}{{\em Phys. Rev. D}
  {\bfseries 106} no.~10, (2022) 103008},
  \href{https://arxiv.org/abs/2204.13781}{{\ttfamily arXiv:2204.13781
  [hep-ex]}}.

\bibitem{Berlin:2019ahk}
A.~Berlin, R.~T. D'Agnolo, S.~A.~R. Ellis, C.~Nantista, J.~Neilson,
  P.~Schuster, S.~Tantawi, N.~Toro, and K.~Zhou, ``{Axion Dark Matter Detection
  by Superconducting Resonant Frequency Conversion},''
  \href{https://dx.doi.org/10.1007/JHEP07(2020)088}{{\em JHEP} {\bfseries 07}
  no.~07, (2020) 088}, \href{https://arxiv.org/abs/1912.11048}{{\ttfamily
  arXiv:1912.11048 [hep-ph]}}.

\bibitem{Berlin:2020vrk}
A.~Berlin, R.~T. D'Agnolo, S.~A.~R. Ellis, and K.~Zhou, ``{Heterodyne broadband
  detection of axion dark matter},''
  \href{https://dx.doi.org/10.1103/PhysRevD.104.L111701}{{\em Phys. Rev. D}
  {\bfseries 104} no.~11, (2021) L111701},
  \href{https://arxiv.org/abs/2007.15656}{{\ttfamily arXiv:2007.15656
  [hep-ph]}}.

\bibitem{Berlin:2022hfx}
A.~Berlin {\em et~al.}, ``{Searches for New Particles, Dark Matter, and
  Gravitational Waves with SRF Cavities},''
  \href{https://arxiv.org/abs/2203.12714}{{\ttfamily arXiv:2203.12714
  [hep-ph]}}.

\bibitem{Ouellet:2018beu}
J.~L. Ouellet {\em et~al.}, ``{First Results from ABRACADABRA-10 cm: A Search
  for Sub-$\mu$eV Axion Dark Matter},''
  \href{https://dx.doi.org/10.1103/PhysRevLett.122.121802}{{\em Phys. Rev.
  Lett.} {\bfseries 122} no.~12, (2019) 121802},
  \href{https://arxiv.org/abs/1810.12257}{{\ttfamily arXiv:1810.12257
  [hep-ex]}}.

\bibitem{Ouellet:2019tlz}
J.~L. Ouellet {\em et~al.}, ``{Design and implementation of the ABRACADABRA-10
  cm axion dark matter search},''
  \href{https://dx.doi.org/10.1103/PhysRevD.99.052012}{{\em Phys. Rev. D}
  {\bfseries 99} no.~5, (2019) 052012},
  \href{https://arxiv.org/abs/1901.10652}{{\ttfamily arXiv:1901.10652
  [physics.ins-det]}}.

\bibitem{Salemi:2021gck}
C.~P. Salemi {\em et~al.}, ``{Search for Low-Mass Axion Dark Matter with
  ABRACADABRA-10~cm},''
  \href{https://dx.doi.org/10.1103/PhysRevLett.127.081801}{{\em Phys. Rev.
  Lett.} {\bfseries 127} no.~8, (2021) 081801},
  \href{https://arxiv.org/abs/2102.06722}{{\ttfamily arXiv:2102.06722
  [hep-ex]}}.

\bibitem{Gramolin:2020ict}
A.~V. Gramolin, D.~Aybas, D.~Johnson, J.~Adam, and A.~O. Sushkov, ``{Search for
  axion-like dark matter with ferromagnets},''
  \href{https://dx.doi.org/10.1038/s41567-020-1006-6}{{\em Nature Phys.}
  {\bfseries 17} no.~1, (2021) 79--84},
  \href{https://arxiv.org/abs/2003.03348}{{\ttfamily arXiv:2003.03348
  [hep-ex]}}.

\bibitem{Boyers:2025qgc}
E.~Boyers, G.~Goldstein, and A.~O. Sushkov, ``{Spin squeezing of macroscopic
  nuclear spin ensembles},''
  \href{https://dx.doi.org/10.1103/PhysRevD.111.052004}{{\em Phys. Rev. D}
  {\bfseries 111} no.~5, (2025) 052004},
  \href{https://arxiv.org/abs/2502.14103}{{\ttfamily arXiv:2502.14103
  [hep-ph]}}.

\bibitem{Foster:2017hbq}
J.~W. Foster, N.~L. Rodd, and B.~R. Safdi, ``{Revealing the Dark Matter Halo
  with Axion Direct Detection},''
  \href{https://dx.doi.org/10.1103/PhysRevD.97.123006}{{\em Phys. Rev. D}
  {\bfseries 97} no.~12, (2018) 123006},
  \href{https://arxiv.org/abs/1711.10489}{{\ttfamily arXiv:1711.10489
  [astro-ph.CO]}}.

\bibitem{Cowan:2010js}
G.~Cowan, K.~Cranmer, E.~Gross, and O.~Vitells, ``{Asymptotic formulae for
  likelihood-based tests of new physics},''
  \href{https://dx.doi.org/10.1140/epjc/s10052-011-1554-0}{{\em Eur. Phys. J.
  C} {\bfseries 71} (2011) 1554},
  \href{https://arxiv.org/abs/1007.1727}{{\ttfamily arXiv:1007.1727
  [physics.data-an]}}. [Erratum: Eur.Phys.J.C 73, 2501 (2013)].

\bibitem{JacksonKimball:2016wzv}
D.~F. Jackson~Kimball, J.~Dudley, Y.~Li, S.~Thulasi, S.~Pustelny, D.~Budker,
  and M.~Zolotorev, ``{Magnetic shielding and exotic spin-dependent
  interactions},'' \href{https://dx.doi.org/10.1103/PhysRevD.94.082005}{{\em
  Phys. Rev. D} {\bfseries 94} no.~8, (2016) 082005},
  \href{https://arxiv.org/abs/1606.00696}{{\ttfamily arXiv:1606.00696
  [physics.ins-det]}}.

\bibitem{Gavilan-Martin:2024nlo}
D.~Gavilan-Martin {\em et~al.}, ``{Searching for dark matter with a 1000 km
  baseline interferometer},''
  \href{https://arxiv.org/abs/2408.02668}{{\ttfamily arXiv:2408.02668
  [hep-ph]}}.

\bibitem{Berlin_2023}
A.~Berlin and K.~Zhou, ``{Discovering QCD-coupled axion dark matter with
  polarization haloscopes},''
  \href{https://dx.doi.org/10.1103/PhysRevD.108.035038}{{\em Phys. Rev. D}
  {\bfseries 108} no.~3, (2023) 035038},
  \href{https://arxiv.org/abs/2209.12901}{{\ttfamily arXiv:2209.12901
  [hep-ph]}}.

\bibitem{Condon}
E.~U. Condon, ``Forced oscillations in cavity resonators,''
  \href{https://dx.doi.org/10.1063/1.1712882}{{\em Journal of Applied Physics}
  {\bfseries 12} no.~2, (1941) 129--132}.

\bibitem{osti_6903398}
W.~B. Smythe, {\em Static and dynamic electricity}.
\newblock New York, NY (USA); Hemisphere Publishing, 1988.

\bibitem{Collin1960FieldTO}
R.~E. Collin, {\em Field theory of guided waves}.
\newblock IEEE Press Series on Electromagnetic Wave Theory. John Wiley \& Sons,
  Nashville, TN, 2~ed., 1990.

\bibitem{Foster:2020fln}
J.~W. Foster, Y.~Kahn, R.~Nguyen, N.~L. Rodd, and B.~R. Safdi, ``{Dark Matter
  Interferometry},'' \href{https://dx.doi.org/10.1103/PhysRevD.103.076018}{{\em
  Phys. Rev. D} {\bfseries 103} no.~7, (2021) 076018},
  \href{https://arxiv.org/abs/2009.14201}{{\ttfamily arXiv:2009.14201
  [hep-ph]}}.

\bibitem{Abragam1983-jt}
A.~Abragam, {\em The principles of nuclear magnetism}.
\newblock International Series of Monographs on Physics. Clarendon Press,
  Oxford, England, 1983.

\bibitem{vankampen}
N.~G. Van~Kampen, {\em Stochastic Processes in Physics and Chemistry}.
\newblock Elsevier, 2007.

\bibitem{Kowalewski2019-wa}
J.~Kowalewski and L.~Maler, {\em Nuclear spin relaxation in liquids}.
\newblock CRC Press, London, England, 2~ed., 2019.

\bibitem{Levitt2008-qj}
M.~H. Levitt, {\em Spin dynamics}.
\newblock Wiley-Blackwell, Hoboken, NJ, 2~ed., 2008.

\bibitem{REDFIELD1965}
A.~Redfield, \href{https://dx.doi.org/10.1016/b978-1-4832-3114-3.50007-6}{{\em
  The Theory of Relaxation Processes}}.
\newblock Elsevier, 1965.

\bibitem{Slichter1990}
C.~P. Slichter, \href{https://dx.doi.org/10.1007/978-3-662-09441-9}{{\em
  Principles of Magnetic Resonance}}.
\newblock Springer Berlin Heidelberg, 1990.

\bibitem{Jackson:1998nia}
J.~D. Jackson, {\em {Classical Electrodynamics}}.
\newblock Wiley, 1998.

\bibitem{Kittel2004}
C.~Kittel, {\em Introduction to Solid State Physics}.
\newblock Wiley, 8~ed., 2004.

\bibitem{SIMPSON201221}
J.~H. Simpson,
  \href{https://dx.doi.org/https://doi.org/10.1016/B978-0-12-384970-0.00002-8}{``Chapter
  2 - instrumental considerations,''} in {\em Organic Structure Determination
  Using 2-D NMR Spectroscopy (Second Edition)}, pp.~21--57.
\newblock Academic Press, Boston, second edition~ed., 2012.

\bibitem{PhysRev.73.679}
N.~Bloembergen, E.~M. Purcell, and R.~V. Pound, ``Relaxation effects in nuclear
  magnetic resonance absorption,''
  \href{https://dx.doi.org/10.1103/PhysRev.73.679}{{\em Phys. Rev.} {\bfseries
  73} no.~7, (1948) 679}.

\bibitem{PhysRev.99.559}
I.~Solomon, ``Relaxation processes in a system of two spins,''
  \href{https://dx.doi.org/10.1103/PhysRev.99.559}{{\em Phys. Rev.} {\bfseries
  99} no.~2, (1955) 559}.

\bibitem{Traficante}
D.~D. Traficante, ``Relaxation. can t2, be longer than t1?''
  \href{https://dx.doi.org/https://doi.org/10.1002/cmr.1820030305}{{\em
  Concepts in Magnetic Resonance} {\bfseries 3} no.~3, 171--177}.

\bibitem{PhysRev.95.8}
N.~Bloembergen and R.~V. Pound, ``Radiation damping in magnetic resonance
  experiments,'' \href{https://dx.doi.org/10.1103/PhysRev.95.8}{{\em Phys.
  Rev.} {\bfseries 95} (1954) 8--12}.

\bibitem{PhysRev.83.34}
H.~B. Callen and T.~A. Welton, ``Irreversibility and generalized noise,''
  \href{https://dx.doi.org/10.1103/PhysRev.83.34}{{\em Phys. Rev.} {\bfseries
  83} (1951) 34--40}.

\bibitem{Kubo_1966}
R.~Kubo, ``The fluctuation-dissipation theorem,''
  \href{https://dx.doi.org/10.1088/0034-4885/29/1/306}{{\em Reports on Progress
  in Physics} {\bfseries 29} no.~1, (1966) 255}.

\bibitem{LANDAU1980333}
L.~D. Landau and E.~M. Lifshitz,
  \href{https://dx.doi.org/https://doi.org/10.1016/B978-0-08-023039-9.50018-4}{``Chapter
  xii - fluctuations,''} in {\em Statistical Physics (Third Edition)}, vol.~5
  of {\em Course of Theoretical Physics}, pp.~333--400.
\newblock Pergamon, 1980.

\end{thebibliography}\endgroup

\clearpage
\newpage
\maketitle
\onecolumngrid
\begin{center}
\textbf{\large Dark Matter Nuclear Magnetic Resonance is Sensitive \\ 
to Dark Photons and the Axion-Photon Coupling} \\
\vspace{0.05in}
{ \it \large Supplemental Material}\\ 
\vspace{0.05in}
{}
{Carl Beadle, Sebastian A. R. Ellis, Jacob M. Leedom, and Nicholas L. Rodd}
\end{center}
\setcounter{equation}{0}
\setcounter{figure}{0}
\setcounter{table}{0}
\setcounter{section}{0}
\renewcommand{\theequation}{S\arabic{equation}}
\renewcommand{\thefigure}{S\arabic{figure}}
\renewcommand{\thetable}{S\arabic{table}}
\renewcommand*{\thesection}{S.\Roman{section}}
\interfootnotelinepenalty=10000 

\setstretch{1.1}

The discussion in this supplemental material is divided into four sections.
First, in Sec.~\ref{ap:conventions} we outline the conventions adopted throughout our work.
In Sec.~\ref{ap:TimeScaling} we explicitly derive the rate at which the DM induced magnetization grows in CASPEr in all relevant regimes.
The essential physics was outlined in the main text and has been seen previously in Ref.~\cite{Dror:2022xpi}.
Given the confusion that has appeared int the literature around these scalings, however, we believe it is useful to outline our additional derivation of these scalings.
From here, in Sec.~\ref{ap:CavityModes} the calculation of the cavity modes used in the main text is detailed.
In Sec.~\ref{ap:MicroToMacroBlochEqns} we provide a detailed description of the transition from a microscopic transition of individual spins to a macroscopic magnetised sample as relevant for DM NMR.
Finally, in Sec.~\ref{ap:Fluctuation}, we derive the power spectral density of spin-projection noise semi-classically as well as quantum mechanically, assuming thermal spin states.

\section{Conventions}
\label{ap:conventions}

We begin by detailing our conventions.
First, we note that care is taken throughout to write the induced magnetic field within the material as $\bH$, whereas the field acting on an individual spin is $\bB$, as $\bH$ is the field induced by currents in the material.
Secondly, when converting to the frequency domain, we find it most convenient to use the following Fourier transform conventions,
\be
\tilde{f}(\w) = \int_{-\infty}^{\infty} dt\, e^{i\w t}\,f(t),
\hspace{0.5cm}
f(t) = \int_{-\infty}^{\infty} \frac{d\w}{2\pi}\,e^{-i\w t}\,\tilde{f}(\w).
\ee
Henceforth, if the bounds are unspecified, integration from $-\infty$ to $+\infty$ is implied.
The time-average of the square of a quantity $f(t)$ can be written in terms of the power spectral density (PSD) as
\be
\langle f^2 \rangle = \lim_{T \to \infty} \frac{1}{2T} \int_{-T}^T dt\, \langle f(t) f^*(t)\rangle = \int \frac{d\w}{2\pi}\, S_f(\w),
\ee
where $2\pi\delta(\w-\w') S_f(\w) \equiv \langle f(\w) f^*(\w')\rangle$ is the PSD of the quantity $f$, and $T$ is the integration time.
It will be useful to have approximate relations for the PSDs in the cases where $f(\w)$ is real, narrow, and peaked at a given frequency $\w_f$, or broad and flat over a bandwidth $\delta \w_f$:
\begin{align}
&S_f(\w)\vert_{\rm narrow} \simeq \pi \langle f^2 \rangle \left[\delta(\w-\w_f) + \delta(\w+\w_f) \right]\!,
\label{eq:fPSDlimits} \\
&S_f(\w)\vert_{\rm broad} \simeq  \frac{\pi\langle f^2\rangle}{\delta \w_f} \Bigg[ \Theta\left(\frac{\delta\w_f}{2}+(\w+\w_f) \right)\Theta\left(\frac{\delta\w_f}{2}-(\w+\w_f)\right)+
\Theta\left(\frac{\delta\w_f}{2}+(\w-\w_f) \right)\Theta\left(\frac{\delta\w_f}{2}-(\w-\w_f) \right)
\Bigg]. \nonumber
\end{align}

The expressions above define the mean value of the PSD, as the power itself will be a stochastic variable if $f(t)$ is a random field.
When computing sensitivities, we assume the PSD of the signal and noise are exponentially distributed to construct the likelihood function as outlined in Refs.~\cite{Foster:2017hbq,Dror:2022xpi}.
In order to determine the expected exclusion sensitivity, we make use of the Asimov procedure~\cite{Cowan:2010js,Foster:2017hbq}, which amounts to setting the data to the mean expected background.
If the mean expected signal and background powers in a bin $k$ are $S_k$, and $B_k$, then the test statistic relevant for setting upper limits is,
\begin{equation}
q = 2 \sum_{k=1}^{N-1} \left[\left(1-\frac{B_k}{S_k + B_k}\right) - \ln\left(1+\frac{S_k}{B_k}\right) \right]\!,
\end{equation}
with the expected 95\% C.L. exclusion being determined by the signal value for which $q=-2.71$.
In the limit where the signal is resolved into many frequency bins (equivalently where the integration time, $T$, is far longer than all other timescales), then the test statistic becomes well approximated by
\begin{equation}
q = - T \int \frac{d\w}{2\pi} \frac{S^2(\w)}{B^2(\w)},
\end{equation}
allowing us to identify $q = -\text{SNR}^2$ to compare with the signal-to-noise ratio (SNR) quantity commonly adopted.

\section{Time-scaling of CASPEr Magnetic Field Signals}
\label{ap:TimeScaling}

We provide here a frequency-domain analysis of the time-scaling of the signal for a generic dark matter-induced magnetic field in an NMR experiment.
The results match those of Ref.~\cite{Dror:2022xpi}, where a similar analysis was performed in the time domain for the axion-gradient signal.
See also the discussion in Ref.~\cite{Zhang_2023} of the time-dependence and optimal scanning strategy for CASPEr-Gradient.

The starting point is the Bloch equations in the $M_\pm \equiv (M_x \pm i M_y)/\sqrt{2}$ basis, given by (cf. Eq.~\eqref{eq:Bloch-pm})
\be
\dot{M}_\pm = - T_2^{-1} M_{\pm} \mp i (\w_0 M_{\pm} - \gamma M_0 H_{\DM}^{\pm}),
\ee
where we have omitted the time-dependence of the relevant quantities for notational simplicity.
Note that the magnetic field induced by the DM generically has both $x$ and $y$ components.
This can be solved straightforwardly as (cf. Eq.~\eqref{eq:Mpsol})
\be
M_\pm(t) = \pm i \gamma M_0\,\int_0^t dt'\, e^{-(t-t')/T_2}e^{\mp i \w_0(t-t')}H_{\DM}^\pm(t')
= \pm i \gamma M_0\,\int_{-\infty}^{\infty}dt'\,W_{\pm}(t,t') H^{\pm}_{\DM}(t').
\label{eq:MpmApp}
\ee
To simplify the subsequent analysis, we have introduced a windowed susceptibility function $W_{\pm}(t,t') = \Theta(t')\Theta(t-t') e^{-(t-t')/T_2} e^{\mp i \omega_0(t-t')}$, with $\Theta(x)$ the Heaviside step-function.
The form of Eq.~\eqref{eq:MpmApp} allows us to define the time-averaged quantity
\be
\langle |M_\pm(t)|^2 \rangle = (\gamma M_0)^2 \int_{-\infty}^{+\infty} \frac{d\w}{2\pi}\, 
\tilde{W}_{\pm}(t,\w)
\tilde{W}^*_{\pm}(t,\w)
S_{\DM}^\pm(\w),
\label{eq:MpmSquared}
\ee
where we introduce the DM PSD as $\langle H_{\DM}^\pm(\w) H_{\DM}^{\pm,*}(\w')\rangle = 2\pi \delta(\w-\w') S_{\DM}^\pm(\w)$.
The quantity $\tilde{W}_{\pm}(t,\w)$ is the Fourier transform of the time-domain function defined above, in particular
\be
\tilde{W}_{\pm}(t,\w)
\tilde{W}^*_{\pm}(t,\w) = \frac{T_2^2}{1+(\omega\pm\omega_0)^2 T_2^2} \left( 1 + e^{-2t/T_2} - 2 e^{-t/T_2} \cos [(\omega \pm \omega_0) t] \right)\!.
\label{eq:W-Wstar}
\ee
We may now use Eq.~\eqref{eq:MpmSquared} to evaluate the left hand side under various assumptions regarding the hierarchies in timescales; in particular, as between transverse relaxation time $T_2$, the DM coherence time $\tau_{\DM}$, and the integration time $T$, effectively the time at which we evaluate the magnetization.
Throughout, we will assume that $\w_0 = m$ in the final results we give, but not in the intermediate steps of the calculation.
We divide the discussion between whether $T_2 \gg T$ or vice versa.

\vspace{0.2cm}
\noindent {\bf Relaxation time $T_2$ exceeds the integration time $T$.}
%
Consider first $T_2 \gg T$.
We study the scaling of the magnetization for different assumptions regarding the hierarchy with $\tau_{\DM}$.
In all cases, we can take the $T_2 \to \infty$ limit of Eq.~\eqref{eq:W-Wstar}, simplifying it to
\be
\lim_{T_2 \to \infty} \tilde{W}_{\pm}(t,\w)
\tilde{W}^*_{\pm}(t,\w) = \frac{2(1-\cos [(\w \pm \w_0) t])}{(\w \pm \w_0)^2}.
\ee
The only difference will be whether we can resolve the signal, determined by the hierarchy between $T$ and $\tau_{\DM}$.
Consider first $T \ll \tau_{\DM}$.
We can then invoke the narrow signal PSD approximation of Eq.~\eqref{eq:fPSDlimits}, writing $S^{\pm}_{\DM}(\w) \simeq \pi |H_\DM^\pm|^2 \left[\delta(\w-m)+\delta(\w+m)\right]$.
We therefore obtain,
\be
\langle |M_\pm(t)|^2 \rangle_{T \ll \tau_{\DM} \ll T_2} \simeq
\langle |M_\pm(t)|^2 \rangle_{T \ll T_2 \ll \tau_{\DM}} \simeq  \frac{1}{2}(\gamma M_0 |H_\DM^\pm|)^2 \,T^2.
\ee
Causality dictates that all results will also have a $\Theta(T)$ present, although we suppress that for simplicity.
The physical origin of the 1/2, which appears in all expressions, is that only half of the DM power is matched with the $\pm \omega_0$ resonant frequency---this is effectively the rotating wave approximation.
If instead we have an integration time $T \gg \tau_{\DM}$, we cannot approximate the magnetic field PSD as a $\delta$-function, and should instead use the broad approximation in Eq.~\eqref{eq:fPSDlimits}, taking the width as $\delta \w = 2\pi/\tau_{\DM}$.
Then we find
\be
\langle |M_\pm(t)|^2 \rangle_{\tau_{\DM} \ll T \ll T_2} \simeq \frac{1}{2} (\gamma M_0 |H_\DM^\pm|)^2\, T \tau_{\DM}.
\ee

\vspace{0.2cm}
\noindent {\bf Integration time $T$ exceeds the relaxation time $T_2$.}
%
If the relaxation time is shorter than the integration time $T$, we must use the full form of the windowed susceptibility given in Eq.~\eqref{eq:W-Wstar}.
However, we can still straightforwardly evaluate the scaling of the signal in the relevant regimes.
If the signal coherence time exceeds $T$, we can once again approximate $S_{\DM}^{\pm}(\w)$ as infinitely narrow, resulting in
\be
\langle |M_\pm(t)|^2 \rangle_{T_2\ll T  \ll \tau_{\DM}} \simeq  \frac{1}{2} (\gamma M_0 |H_\DM^\pm|)^2\, T_2^2.
\ee
The other cases require $T \gg \tau_{\DM}$, so that the DM signal is well resolved and can be approximated as broad.
Performing the integrals and carefully treating the hierarchy between $T_2$ and $\tau_{\DM}$, we find,
\be
\langle |M_\pm(t)|^2 \rangle_{T_2\ll \tau_{\DM}\ll T} \!&\simeq&\! \frac{1}{2} (\gamma M_0 |H_\DM^\pm|)^2\,T_2^2, \\
\langle |M_\pm(t)|^2 \rangle_{\tau_{\DM}\ll T_2\ll  T} \!&\simeq&\! \frac{1}{2} ( \gamma M_0 |H_\DM^\pm|)^2\, T_2 \tau_{\DM}.
\ee

\section{Driven cavity modes}
\label{ap:CavityModes}

Here we elaborate on the cavity modes that were discussed in the main body of the text.
A quick review of cavity modes is provided before a discussion of the selection rules for a DM signal that are dictated by the cavity's geometry.

\subsection{Cavity Basics}

We specialise the analysis to cylindrical cavities as the shielding of the CASPEr-Gradient experiment is well approximated by a cylinder, and because of its relative simplicity.
All of our qualitative findings will extend to more general geometries.
The cavity modes are determined by solving source-less Maxwell equations.
We align the bore of the cylinder with the $z$-direction and decompose the electric and magnetic field as follows,
\be
\begin{Bmatrix} \bE(\bx, t) \\ \bH(\bx, t) \end{Bmatrix} 
= \int \frac{d \w dk}{(2\pi)^2}  \,  \begin{Bmatrix} \bE (x,y,k,\w) \\ \bH(x,y,k,\w) \end{Bmatrix}  \, e^{i (k z- \w t)}.
\label{apeq:FieldFourierDecomposition}
\ee
The fields satisfy the wave equation
\be
\left[ \nabla_T^2 + \mu \varepsilon \, \w^2 - k^2 \right] \,  \begin{Bmatrix} \bE(x,y,k,\w) \\ \bH(x,y,k,\w) \end{Bmatrix} 
= \mathbf{0},
\label{apeq:ModeWaveEquation}
\ee
where $T$ indicates a sum over the transverse directions, whilst $\mu$ and $\epsilon$ are the permeability and permittivity.

The cavity walls are formed by the magnetic shielding which is used is to mitigate the effects of low-frequency external magnetic fields.
This is normally achieved by using a metal alloy with very high magnetic permeability whose effect on spin-dependent interactions with the sample has been considered in Refs.~\cite{JacksonKimball:2016wzv,Gavilan-Martin:2024nlo}.
These alloys usually have electrical conductivity values somewhere in the range $10^6 - 10^8$\,S/m, typical of metallic materials.
For frequencies away from a cavity resonance, it is then sufficient to approximate the walls of the cavity as perfectly conductive, recovering the following simple boundary conditions:
\be
\bn \times \bE = \mathbf{0},
\hspace{0.5cm}
\bn \cdot \bH = 0,
\label{apeq:BoundaryConditionsPerfectConductor}
\ee
where $\bn$ denotes the normal direction from the surface of the conductor.
For finite conductivity, the analysis is modified by introducing the generalised Ohm's law contribution to the quality factor.
In general, the presence of the CASPEr spin sample within the cavity will provide a small modification to the modes.
For simplicity, we neglect the impact of the sample on the calculation as our projections will only be minorly impacted.

In order to impose the boundary conditions, we label the two end-caps of the cylinder by $\mathcal{C}$ and the circular wall by $\mathcal{S}$.
The cylinder is taken to have a radius $R$ and height $l_z$.
Imposing the boundary conditions of Eq.~\eqref{apeq:BoundaryConditionsPerfectConductor} on $\mathcal{C}$ determines the allowed values of $k$, and imposing them on $\mathcal{S}$ allows us to split the possible field configurations into two groups, the transverse magnetic (TM) and transverse electric (TE) modes.
The TM modes satisfy $E_z \big|_{\mathcal{S}} = 0$ with $H_z = 0$ throughout the cavity.
The TE modes instead satisfy $\pd_n H_z \big|_{\mathcal{S}} = 0$ and $E_z = 0$.
The two sets of modes are further labelled by integers $\ell = (m,n,p)$ that determine the mode number in the azimuthal, radial, and longitudinal direction.

\begin{figure}[!t]
\centering
\includegraphics[width=0.8\linewidth]{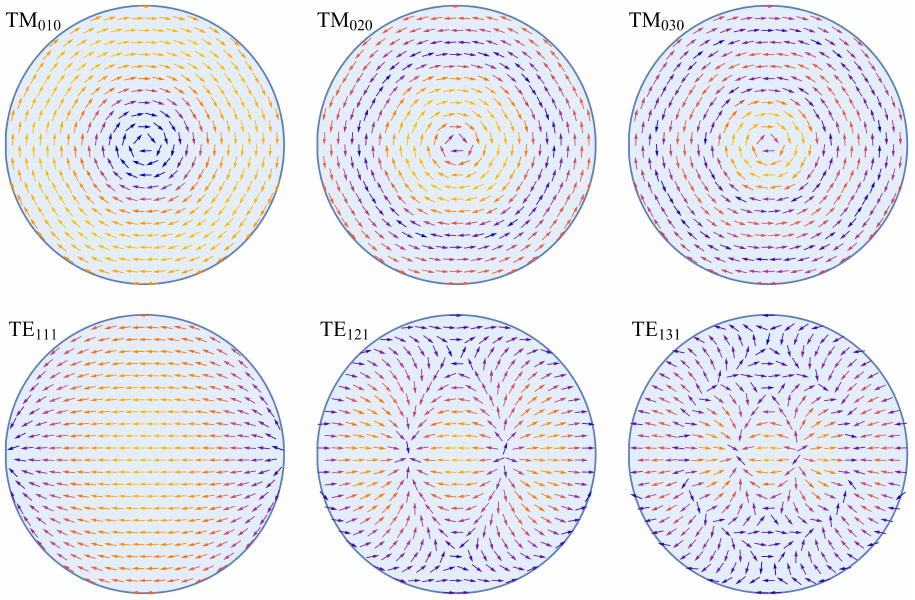}
\caption{The transverse magnetic field geometry of the lowest lying modes of a cylindrical cavity.
We set $\mu = \varepsilon = 1$ and show a section at $z=0$ (i.e. at a cap).
The heat map is such that yellow corresponds to larger field values and blue smaller.
}
\label{fig:LowCylindricalModes}
\end{figure}

With the above requirements, the TM modes are given by
\begin{equation}\begin{aligned}
E_z(\bx,t) &= \phi(x,y) \, \cos(p \pi z/l_z), \\
\bE_T(\bx,t) &= \nabla_T \phi(x,y)  \, \frac{\left( p \pi/l_z \right) }{\left( p \pi / l_z \right)^2 - \varepsilon \mu \, \w^2} \sin(p \pi z/l_z), \\
\bH_T(\bx,t) &=  - \hz \times \nabla_T \phi(x,y) \, \frac{i \, \varepsilon \, \w }{\left( p \pi /l_z \right)^2 - \varepsilon \mu \, \w^2} \cos(p \pi z/l_z),
\label{apeq:TMModesEz}
\end{aligned}\end{equation}
where $\phi$ is a function satisfying the two dimensional wave equation and $\phi \big|_{\mathcal{S}} = 0$.
The TE modes are similarly described
\begin{equation}\begin{aligned}
H_z(\bx,t) &= \psi (x,y) \, \sin(p \pi z/l_z), \\
\bH_T(\bx,t) &=  \nabla_T \psi(x,y) \, \frac{\left( p \pi / l_z \right)}{\left( p \pi / l_z \right)^2 - \varepsilon \mu \, \w^2} \cos(p \pi z/l_z),\\
\bE_T(\bx,t) &= \hz \times \nabla_T \psi(x,y) \, \frac{i \, \mu \, \w  }{\left( p \pi / l_z \right)^2 - \varepsilon \mu \, \w^2} \sin(p \pi z/l_z).
\label{apeq:TEModes}
\end{aligned}\end{equation}
Here $\psi$ plays the role of $\phi$, and satisfies the modified boundary condition $\pd_n \psi \big|_{\mathcal{S}} = 0$.
Note for the TM modes $p$ begins at 0, whereas for the TE modes the smallest value is 1.
The transverse fields are described by Bessel functions,
\be
\phi(x,y) \propto J_m(j_{mn} r/R) \, e^{im\theta},
\hspace{0.5cm}
\psi(x,y) \propto J_m(j_{mn}^{\prime} r/R) \, e^{im\theta}.
\label{apeq:phipsi}
\ee
Here $r$ and $\theta$ are the cylindrical coordinates, $j_{mn}$ denotes the $n$th zero of the $m$th Bessel function and $j_{mn}^{\prime}$ denotes the  $n$th zero of the $m$th Bessel function's derivative.
The full set of modes form an eigenbasis of functions and so are orthogonal to each other when integrating over the volume of the cavity.
The constants of proportionality in Eq.~\eqref{apeq:phipsi} can be chosen to ensure the modes are orthonormal, for instance we assume $\int dV\,\bE_{\ell} \cdot \bE^*_{\ell'} = \delta_{\ell \ell'}$, where the integral is taken over the shielded volume, $V$.
The eigenfrequencies are
\be
\w_{\ell}^2 = \frac{1}{\varepsilon \mu} \left[ (j^{\left( \prime \right)}_{mn}/R)^2 + (\pi \, p/l_z)^2 \right]\!.
\label{eq:ModeFrequencies}
\ee
The geometry of the magnetic field for the lowest lying cavity modes is shown in Fig.~\ref{fig:LowCylindricalModes} (cf. Fig.~\ref{fig:Placement}).

\subsection{Driven Cavity Modes}

The above discussion provides us with a complete set of modes within the cavity, allowing for a decomposition of arbitrary electric and magnetic fields.
These can then be used to determine the form of the fields generated by an effective charge or current configuration as,
\be
\left( \nabla^2 - \varepsilon \mu \, \pd_t^2 \right) \begin{Bmatrix} \bE (\bx,t) \\ \bH (\bx,t) \end{Bmatrix} = \begin{Bmatrix}
\frac{1}{\varepsilon} \left( \nabla \rho_{\text{eff}} + \varepsilon \mu \, \pd_t \bJ_{\text{eff}} \right)\\ - \nabla \times \bJ_{\text{eff}}
\end{Bmatrix}\!.
\label{apeq:WaveEqnsDriven}
\ee
The electric field can be decomposed as,
\be
\bE(\bx,t) = \int \frac{d \w}{2 \pi} \, e^{-i\w t} \, \sum_{\ell} c_{\ell}(\w) \, \bE_{\ell}(\bx).
\ee
This result can be applied for the TE and TM modes separately.
The coefficients are determined from Eq.~\eqref{apeq:WaveEqnsDriven} as,
\be
c_{\ell}(\w) = \frac{i \w}{\varepsilon(\w_{\ell}^2-\w^2)}\int dV\, \bJ_{\rm eff}\cdot \bE^*_{\ell}.
\label{apeq:cell}
\ee
The integral in this expression dictates a set of selection rules relating the form of the DM current to the modes generated, as explored in Eq.~\eqref{eq:Jeff-coup} in the main text.
Note that the gradient of the charge density has no effect on the divergence free field modes we consider~\cite{Berlin_2023,Condon,osti_6903398,Collin1960FieldTO}.
The magnetic fields are fixed from the above discussion through Faraday's law.
In detail,
\be
\bH(\bx,t) = \int \frac{d \w}{2 \pi} \, e^{-i\w t} \sum_{\ell} \frac{\w_{\ell}}{\w} \, c_{\ell}(\w) \, \bH_{\ell}(\bx).
\ee
The focus of CASPEr-Gradient is on the DM parameter space where $\omega \ll \omega_{\ell}$.
Therefore the induced magnetic field is parametrically larger than the induced electric field, as explained in the main text.

Given DM is non-relativistic and has a characteristic speed of $v \sim 10^{-3}$, the spatial scale the DM varies over is set by the de Broglie wavelength, $\lambda_{\rm dB} \sim 1/mv \sim 1\,{\rm km}\,(1\,\mu{\rm eV}/m)$
(see e.g. Ref.~\cite{Foster:2020fln} although the more accurate scale is the coherence length~\cite{Cheong:2024ose}), which is far larger than any relevant experimental scale.
This implies that it is an excellent approximation to take the DM effective current in Eq.~\eqref{apeq:cell} as spatially uniform, so that $c_{\ell} \propto \bJ_{\rm eff} \cdot \int dV\,\bE^*_{\ell}$, as in Eq.~\eqref{eq:Jeff-coup}.
The symmetry of the modes is such that for the TE modes we require $m=1$, $p$ odd, and $n$ arbitrary; $\int dV\,\bE^*_{\ell}$ is then purely transverse, implying it is accessible to the dark-photon current, but not the axion-photon coupling.
For the TM modes, $m=p=0$, and the integral can be evaluated as
\be
{\rm TM\,modes\!:}\hspace{0.2cm} \int dV\,\bE_{0n0}^* = (-1)^{n+1} \frac{2}{j_{0n}} \sqrt{V}\, \hz.
\ee
Being purely in the $z$-direction, all DM currents can excite these modes.

As long as the DM frequency is well below $\w_{\ell}$, the lowest modes will be dominantly excited: higher modes are always more oscillatory and therefore suppressed by the volume integral, additionally as the excited modes have an even larger frequency, the DM is further still off resonance.
The key modes are therefore TM$_{010}$ and TE$_{111}$.
If the ratio $R/l_z > 0.492$, then $\w^{\text{TE}}_{111} > \w^{\text{TM}}_{010}$ and vice versa.
Therefore, if the diameter of the cylinder is roughly the same size as its length these frequencies are comparable.

An important point to emphasise is that the modes are not spatially uniform throughout the cavity.
This determines the magnetic field as seen by the sample.
The relative sizes of the sample and shielding matter; a small sample will effectively see a uniform field across it whereas a large sample will resolve the mode.
If the spin sample were comparable in size to the shielding, then the average magnetic field across it will be zero.
In reality, one could improve upon this because the exact spatial profiles of all of the modes are known, meaning that even though the net field is zero, the magnetic field in a given domain of the sample is known. 
This allows one to either take local readings of the magnetic flux with a magnetometer across a portion of the sample, or to form particular geometries in the pickup loops such that the flux coming from different parts of the sample is additive.
For example, the TE$_{111}$ mode has a cross-section of the form shown in Fig. \ref{fig:LowCylindricalModes} near the upper cap.
The magnetic field for this mode varies as $\cos(\pi z/l_z)$, such that at the bottom cap the magnetic field lines point the opposite direction. 
A gradiometer configuration of the pickup loop could exploit this variation, such that the two contributions are additive and the field can be seen.

Nevertheless, given the relatively small size of the CASPEr-Gradient sample compared to the shielded region, the field will be close to uniform across it.
The magnetic field of the TM modes varies only radially, proportional to $J_1(j_{01} r/R)$.
The field is largest at $r = j'_{11}/j_{01} R \simeq 0.77 R$, and for a small deviation $\delta r$ away from value we have,
\be
\frac{\delta H}{H_{\text{max}}} = - \frac{1}{2} \left( \frac{j_{01}}{j'_{11}} \right)^2 \left[ \left( j'_{11} \right)^2 - 1 \right] \left( \frac{\delta r}{R} \right)^2\!.
\ee
For the sample, that implies the field varies by $\simeq 6\%$ across the sample (cf. Fig.~\ref{fig:Placement}).
The analogous result for the TE modes is that the field takes its maximum values at the centre of the shielding at $r=0$, but near the caps $z = 0$ or $l_z$.
The fractional size of the magnetic field away from the optimal placement is set by
\be
\frac{\delta H}{ H_{\text{max}}} \simeq - \frac{1}{2} \left(\frac{2+\cos 2 \theta }{4} \right) \left( j'_{11} \right)^2 \left( \frac{\delta r}{R} \right)^2 - \frac{\pi^2}{2} \left( \frac{\delta z}{l_z} \right)^2\!.
\ee

\section{Microscopic to macroscopic description}
\label{ap:MicroToMacroBlochEqns}

The Bloch equations given in Eq.~\eqref{eq:bloch} are often justified heuristically as a generalization of the evolution of a single spin in a magnetic field.
The equations can, however, be derived from a microscopic perspective~\cite{Abragam1983-jt,vankampen, Kowalewski2019-wa,Levitt2008-qj}. 
Doing so clarifies the microphysical origin of the macroscopic equations.
In this appendix, we discuss some of the salient aspects of this microphysical description and how they affect the analysis we present in the main text.
Throughout this appendix we reserve the symbol $H$ for the Hamiltonian and use $B$ for magnetic fields.

\subsection{Master relaxation equation}

A quantum spin system can be described by a density operator $\rho = p_j\ket{\psi_j}\bra{\psi_j}$, where $p_j$ is a weight for the state labelled by $\psi_j$, and summation is implied. 
This density matrix evolves according to the Liouville-von Neumann equation:
\be
\frac{d}{dt} \rho = i \left[ \rho, H \right]\!.
\label{apeq:LiouvillevonNeumann}
\ee
Observables can then be calculated by taking the weighted trace $\langle Q \rangle = \text{tr} \left( \rho \, Q \right)$.
We may decompose the Hamiltonian into time independent and time dependent parts, $H = H_0 + H_I(t)$.
This choice allows us to account for interactions between the spin system and the reservoir, which is treated as being classical, in a simple manner.
In the interaction picture, a quantity $X$ is transformed as $\bar{X} \equiv \exp \left( i H_0 t \right) X \exp \left( - i H_0 t \right) $, and the evolution of the spin density per Eq.~\eqref{apeq:LiouvillevonNeumann} is therefore
\be
\frac{d}{dt} \bar{\rho} = i \left[ \bar{\rho}(t), \bar{H}_I(t) \right]\!.
\ee
We assume that environmental processes are characterised by a short correlation time $\tau_c$ such that $\tau_c \, \langle \bar{H}_I^2\rangle^{1/2} \ll 1$.
This corresponds to assuming that the environment is a thermal system, and $\tau_c \ll t \ll \langle \bar{H}_I^2\rangle^{-1/2}$.
Here $\langle \bar{H}_I^2\rangle^{1/2}$ is the root-mean-square energy driving the system from the environment which dictates a natural largest timescale after which the system will have fully relaxed.
Correlations between the density matrix and the interaction Hamiltonian representing the environment can therefore be neglected.
We further assume that the evolution is only dependent on the current state of the system, meaning that we take the Born-Markov approximation.
This allows us to rewrite the evolution equation in the form of the master equation~\cite{REDFIELD1965,Slichter1990},
\be
\frac{d}{d t} \bar{\rho}(t) = - \int_0^{\infty} d\tau \left[ \bar{H}_I(t),\,\left[ \bar{H}_I(t + \tau), \bar{\rho} \left( t \right) \right]  \right]\!.
\label{apeq:MasterEquationInt}
\ee
The upper limit of the integral has been extended to infinity as the short correlation times ensure that the integrand does not have support at large times.
Formally, the calculation accounts for semi-classical fluctuations about a background density operator $\rho_0$ in thermal equilibrium, so the density matrix should be modified to $\rho \to \rho-\rho_0$.
The classical background magnetisation $M_0$ can be thought of as being represented through the thermal background $\rho_0$, as we will see upon derivation of the Bloch equations.

\subsection{The interaction Hamiltonian}

The non-relativistic limit of the Dirac Lagrangian leads to the appropriate description for the interaction between the fermion spins and EM fields. 
The leading interaction between a magnetic field and spin in the $v \ll 1$ expansion is given by 
\be
H_n \supset - \gamma \, \bB  \cdot  \mathbf{s}_n,
\label{apeq:Hamiltonian}
\ee
with the index $n$ indicating that the full Hamiltonian is the sum over all spins.
The coefficient $\gamma$ is the species-dependent gyromagnetic ratio that accounts for substructure of the nuclei.
The Hamiltonian of Eq.~\eqref{apeq:Hamiltonian} is valid at the location of the spin in question, $\mathbf{s}_n$.
To remove any ambiguities that may arise when taking volume averages of quantities to obtain macroscopic quantities, we can define the spin-density operator
\be
\mathbf{S}(\bx) = \sum_n\mathbf{s}_n \, \delta(\bx - \bx_n),
\ee
such that $\int dV \, \mathbf{S} = \sum_n \mathbf{s}_n$.
The validity of this definition is ensured as long as there is a sufficiently large scale separation between the typical inter-spin distance and the typical scale of variation of the applied magnetic field. %
This is always the case for the situations we consider.
We can then decompose the applied magnetic field into a background field $\mathbf{B}_0$ entering the free Hamiltonian $H_0$, and interacting fields $\mathbf{B}_I$ in the interaction Hamiltonian $H_I$, with $\mathbf{B} \to \mathbf{B}_0 + \mathbf{B}_I$ in Eq.~\eqref{apeq:Hamiltonian}.

\subsection{Thermal magnetisation vector: \texorpdfstring{$M_0$}{M0}}

To derive the Bloch equations, we first define the macroscopic magnetisation vector in terms of the spin-density operator and a spatial window function $f(\bx)$ normalised so that $\int dV f(\bx) = 1$~\cite{Jackson:1998nia}
\be
M^a(\bx, t) \equiv \gamma \, \int d^3 \bx' \, f (\bx') \, \langle S^a ( \bx - \bx' , t) \rangle,
\label{apeq:MagnetisationVector}
\ee
where $\langle\cdot\rangle$ indicates that the weighted trace of the spin-density operator is taken, and $a$ is a spatial index.
The precise expression for the window function $f$ is left unspecified here.
However, it is included so as to allow future averaging of local spin effects over macroscopic length scales, resulting in equations containing quantities which are relevant on these scales.

For the purposes of the experiment, the $M_0$ in the Bloch equations corresponds to a volume averaging across the entire spin sample, such that any local inhomogeneities are integrated out.
Assuming an external magnetic field applied along the $z$-direction, the resulting $\bM_0 = M_0 \hz$.
Starting from the thermal density matrix $\rho_0$, weighted by thermal Boltzmann factors, one can obtain Curie's law for the magnetisation vector from Eq.~\eqref{apeq:MagnetisationVector} as in Ref.~\cite{Kittel2004}
\be
M_0 =  n \gamma J \, \mathcal{B}_J \!\left( \frac{\gamma \, B_0}{ T_s} \right) \! ,
\label{apeq:CuriesLaw}
\ee
where $n$ is the number density of spins, $J$ is the largest spin value, $k_B$ is Boltzmann's constant, and $T_s$ is the spin temperature.
The function $\mathcal{B}_J$ is called the Brillouin function, taking the form
\be
\mathcal{B}_J(x) = \frac{1}{J} \left[ \left( J +\frac{1}{2} \right)\, \coth\! \left( \left( J +\frac{1}{2} \right) x \right) - \frac{1}{2} \coth\! \left( \frac{x}{2}\right)\right]\!.
\label{eq:BrillouinFunction}
\ee
When $J=1/2$, the result reduces to
\be
M_0 = \frac{n \gamma}{2} \tanh\! \left( \frac{\gamma B_0}{2 T_s} \right)\!.
\ee

\subsection{Non-thermal magnetisation vector: \texorpdfstring{$M^i$}{Mi}}

The full interaction Hamiltonian contains the interactions of spins with the magnetic field, the lattice and other spins.
However, the analysis can be simplified by treating all interactions as between a spin and a magnetic field, with the latter being generated by different sources.
Therefore we decompose the $B_I$ into contributions from excitations of the effective cavity containing the spin sample, fields generated by the spins, inhomogeneities in the applied magnetic field $B_0$, and the back-reaction field arising from the readout.
The volume average of inhomogeneous fields is assumed to be zero.
In the case of spin-induced fields, this is a good approximation.
We can then use Eq.~\eqref{apeq:MasterEquationInt}, expanded as
\be
\dot{\rho} = i \, \left[ \rho , H_0 + H_I (t) \right] - e^{-i H_0 t} \int_0^{\infty} d\tau \left[  \bar{H}_I(t), \left[ \bar{H}_I(t + \tau),\, \bar{\rho}(t) \right]  \right]e^{i H_0 t}, 
\label{apeq:rhoEoM}
\ee
to obtain the equation of motion for the magnetisation.
Retaining only the first term above, we recover the dissipationless part of the Bloch equations,
\be
\dot{M}^a(\bx, t) = \gamma^2 \, \epsilon^{abc} \, \sum_n  f(\bx - \bx_n) \langle \, \mathbf{s}_n^b \, \rangle \, B^c( \bx_n , t) \simeq \gamma \, \epsilon^{abc} \,  M^b(\bx, t) \, B^c(\bx,t),
\ee
where latin indices indicate spatial components of the vectors.
To obtain the final expression above, we make the approximation that the magnetic field does not vary significantly over the scale of the window function $f(\bx)$.
Since the characteristic scale of this function is the typical inter-molecular spacing of $\mathcal{O}(\text{nm})$, this is a reasonable assumption.
The magnetic field $B^c(\bx,t)$ above includes the applied background field $B_0$, as well as the signal and radiation damping fields.
It is assumed that shimming leads to inhomogeneities whose volume average is negligibly small~\cite{SIMPSON201221}.

Dissipation in the system can be recovered by retaining the higher-order term in Eq.~\eqref{apeq:rhoEoM}.
This gives rise to contributions that are typically of the form
\be
\dot{M}^a \sim \gamma^3 \, \int d^3x' \int_0^{\infty} d \tau  \,f( \bx') \, R^{abcd}\, \langle \, S^b(\bx-\bx')\, \rangle \, B^c(t) \, B^d(t + \tau),
\label{apeq:Schematic}
\ee
where $R^{abcd}$ is a tensor contracting the indices of spin and magnetic fields with terms that can also have a trigonometric time dependence.
The result is that dissipation depends on the auto-correlation function of the magnetic fields acting on the spin system.
For temporally and spatially inhomogeneous fields these will typically be non-zero.
Summing over all such contributions of the form given in Eq.~\eqref{apeq:Schematic} leads to spin relaxation. 
Therefore, they dictate the magnitude of $T_1$ and $T_2$ in the Bloch equations
\be
\dot{\bM} = \bM \times \gamma \mathbf{B} - \frac{M_x \hx+M_y \hy}{T_2} - \frac{(M_z-M_0)\hz}{T_1}.
\label{apeq:BlochEqs}
\ee
We now explain the significant processes that set the relevant timescales.

\subsection{Timescales: \texorpdfstring{$T_1$}{T1}, \texorpdfstring{$T_2$}{T2}, and \texorpdfstring{$T_2^{*}$}{T2s}}

An intuitive model of both $T_1$ and $T_2$ relaxation timescales can be found in early works on NMR~\cite{PhysRev.73.679,PhysRev.99.559}.
From the form of the Bloch equations in Eq.~\eqref{apeq:BlochEqs}, we can see that the organising principle behind relaxation is to determine whether a process contributes to longitudinal relaxation along $\bM_0$ or to transverse relaxation along the two perpendicular directions.
An observation that can be made is that the magnitude of the magnetisation vector in any direction at any time must remain $|M^a| \leq M_0$, if prepared such that it is initially so.
This requirement can be used to show that $T_2 < 2 \, T_1$~\cite{Traficante}.
Different microscopic processes contribute to the two timescales; a subset of interactions that need to be accounted for are spin-lattice, spin-spin, and back-reaction.

Consider the rate of transition $\Gamma_{\uparrow \downarrow}$ from the state $\ket{\uparrow}$ to the state $\ket{\downarrow}$, which corresponds to the timescale $T_1$.
By Fermi's golden rule, it is proportional to the square of the matrix element: $\bra{\downarrow} H_I \ket{\uparrow}$.
Using the interaction Hamiltonian in Eq.~\eqref{apeq:Hamiltonian}, we see that only magnetic fields in the transverse directions have non-vanishing matrix elements, because the states are orthonormal eigenstates of the spin-$z$ operator.
Therefore, we can estimate the mean transition rate as
\be
\Gamma_{\uparrow \downarrow} \sim \gamma^2 \langle B_T^2 \rangle\, \mathcal{S}(\w_0),
\label{apeq:RateT1}
\ee
where $(\gamma B)^2$ arises from the state transition matrix element and $\mathcal{S}(\w_0)$ carries the spectral information of the magnetic fields evaluated at the Larmor frequency $\w_0$.
Some of these magnetic fields are the result of spin-lattice interactions, where a spin exchanges energy with the lattice, contributing to relaxation in all directions.
The energy exchange with the thermal bath of the lattice allows a spin flip that would otherwise be forbidden in a fixed background magnetic field.
(That only transverse magnetic field fluctuations $B_T$ can cause $T_1$ relaxation can in principle also be seen from the explicit form of Eq.~\eqref{apeq:Schematic}, although as the expression is quite involved we do not state it here.) 
These transitions are conceptually the same as those in the two transverse directions, i.e. from $\ket{s_x; +}$ to $\ket{s_x; -}$ or from $\ket{s_y; +}$ to $\ket{s_y; -}$.
These transitions describe the same kind of dissipative spin flips in the transverse directions and will also have rates of the form given in Eq.~\eqref{apeq:RateT1}.
Alternatively, consider the transition from $\ket{s_x; +}$ to $\ket{s_y; \pm}$, related to the rotation of a spin.
One can see that the corresponding matrix elements for the process do not vanish on insertion of any spin operator, meaning that this transition can occur from magnetic fields in any direction.
These transitions are related to the dephasing of the spins, and can occur for arbitrary mixed states, not only eigenstates of the spin operators.
We see from this that most processes generally contribute to $T_2$, while only a subset contribute to $T_1$.
Sufficiently long $T_1$ has been achieved experimentally so that in practice, $T_2$ is the limiting quantity for DM searches.

Spin-spin interactions contribute largely to $T_2$.
These interactions are the result of local magnetic fields generated by each of the magnetic dipoles $\mathbf{m}$,
\be
\bB(\bx) = \frac{1}{4\pi} \left[ \frac{3 \hx \left(\hx \cdot \mathbf{m}\right) - \mathbf{m} }{\left|\bx \right|^3}\right] \!.
\label{apeq:LocalDipoleField}
\ee
A na\"ive estimate of the timescale associated to these local magnetic fields can be derived by taking $\mathbf{m} \sim \gamma$ and $x \sim n^{-1/3}$, so that nearby spins experience an additional field of magnitude $\gamma n$, and therefore precess and decohere on a timescale $T_2 \sim 1/n\gamma^2$.
This estimate is confirmed through the more general analysis above.
Inserting the magnitude of the local magnetic fields into Eq.~\eqref{apeq:Schematic}, we find the same result.
Intuitively the volume integral over the spin-density will become the magnetisation vector $M$ as per the definition, the magnetic fields have values of size $\gamma n$ and their integral will be coherent only over the timescale $1/(n \gamma^2)$.
Combining these we see ~$\dot{M} \sim \gamma^2 \, M \, (n \gamma )^2 \, \{ 1 / (n \gamma^2 )\} = M ( n \gamma^2 )$, recovering the estimate for $T_2$.

It is common practice in the literature to account for effects such as detector back-reaction and magnetic field inhomogeneities by introducing the timescale $T_2^*$.
Where $T_2$ may be thought of as the parameter coming from the microphysics of the sample itself, $T_2^*$ is determined by both the microphysics of the sample \emph{and} interactions between the spins and detection apparatus.
$T_2$ can be derived if a complete microscopic description of the spins is known.
In practice we do not do this, although the size of the parameter can be determined to the order of magnitude level as above.
From that estimate, the expectation is that $T_2 \sim \text{ms}$ for $^{129}\text{Xe}$ and $T_2 \sim 10\,\mu\text{s}$ for $^3\text{He}$.

The definition of $T_2^*$, arising as a sum of dissipation from within and outside the sample, implies it can be decomposed as

\be
\frac{1}{T_2^*} \equiv \frac{1}{T_2} + \frac{1}{T_b},
\ee
where $T_b$ is a timescale associated with spin-bath interactions.
(Here the bath captures the broader environmental effects that are not included in the previous spin-spin analysis.)
Thus, it must always satisfy $T_2^* \leq T_2$. 
A dominant contribution to $T_2^*$ comes from the small inhomogeneities in the externally applied magnetic field.
This induces a spread in the Larmor frequencies of the spins across the sample, causing them to dephase.
The stated level of inhomogeneities in the field is $2$ ppm, which can be translated directly to the spread in frequencies \cite{Walter:2023wto}.
For the mass scales of interest we see that this implies that $T_2^*$ can be estimated to be $0.1$  milliseconds.

Radiation damping occurs from back-reaction of the sample against the pickup coils, by Faraday's law of induction.
Not all of the magnetic fields from the coils can be accounted for in $T_2^*$;  some must be included directly in Bloch's equations.
This is because a component will be spatially coherent across the entire spin sample and so must enter the Bloch equations as $\bB$ to be self-consistent.
The effect of radiation damping can be estimated by using considerations of the torque on the magnetisation vector as was done in a very early analysis of Ref.~\cite{PhysRev.95.8}.
There, it was shown that one must solve a coupled system of equations, whereby the dynamics of the magnetic field are accounted for in the circuit equations and the spins by the usual Bloch equations.

\section{Fluctuation-dissipation theorem for Spin Projection Noise}
\label{ap:Fluctuation}

The fluctuation-dissipation theorem \cite{PhysRev.83.34,Kubo_1966} allows us to determine the thermodynamic fluctuations of the spins from the susceptibility of the system.
Here we follow the approach to dissipation taken in Ref.~\cite{LANDAU1980333}, but applied to the system of spins.
Semi-classically, we can compute the total energy dissipated by the system up to a time $T$ as,
\be\begin{aligned}
Q &= \int_{-\infty}^T dt \,\overline{\frac{dE}{dt}} = \int_{-\infty}^T dt \left\langle\frac{\pd H_{I}}{\pd B_n^a}\right\rangle \pd_t B_n^a \\
&= - \gamma \int_{-\infty}^T dt\,\sum_n \langle s_n^a\rangle \, \pd_t B_n^a.
\end{aligned}\ee
In the first line, we have used that the only time-dependence on the average energy $\overline{E}$ arises through $B_n^a$, and used the Hellmann-Feynman theorem to relate $\overline{\pd_B E} = \Big\langle\pd_B H_{I}\Big\rangle$. 
When writing $\langle{\pd_BH_{I}}\rangle$, we are implicitly assuming that the averaging includes the quantum averaging over states and the statistical averaging.
To get the second line above, we have used the form of the interaction Hamiltonian in Eq.~\eqref{apeq:Hamiltonian}.

Next, we take the system to be a discrete number of spins, interacting with a magnetic field.
Their time-evolution is dictated by the interaction Hamiltonian of Eq.~\eqref{apeq:Hamiltonian}.
Additionally, we define the susceptibility of a single spin through the following integral, to encode causal dynamics of the spins
\be
\gamma\, \langle s_n^a \left(t \right) \rangle = \int d \tau  \, \alpha_{ab}(\tau) \, B^b_n( t - \tau) ,
\label{apeq:SuscepDef}
\ee
where here we sum over repeated indices $a,b$ that indicate spatial components, while the index $n$ identifies the spin, and the position at which the magnetic field is evaluated.
Note that $\alpha$ must be proportional to a Heaviside function to have the correct causal properties.
In the frequency domain, the above expression can then be written as
\be
\gamma \langle s^a_n(\w) \rangle = \alpha_{ab}(\w) \,B^b_n(\w).
\ee
We can use this definition to evaluate $Q$, which after symmetrisation leads to
\be
Q = \frac{i}{2}\sum_n \int \frac{d \w}{2 \pi } \, \w B_{n}^a( \w)\, B_{n}^b(-\w) \left[\alpha_{ab}^*(\w) -\alpha_{ba}(\w) \right]\!,
\label{apeq:DissRate}
\ee
where the ordering of the indices $a,b$ is physical, as $\alpha_{ab}(\w)$ is not necessarily a symmetric matrix.
To obtain this result we took $T \to \infty$ implying that this is a steady state calculation of the noise.

From here, the aim is to use Eq.~\eqref{apeq:DissRate} to obtain the PSD of the spins.
We can achieve this by mapping the result onto a dissipation rate that is directly dependent on the spin PSD.
On general grounds, we would expect that the time-averaged rate of change of energy for the system should scale as
\be
Q \sim \Delta \w  \,\overline{S_{s^a_n s^b_m}}(\w) \,\mathcal{R}(\w;a,b;n,m) \ .
\label{apeq:RoughQ}
\ee
This is simply the statement that the thermally-averaged spin state PSD, $\overline{S_{s^a_n s^b_m}}(\w)$, should set the average rate at which spins dissipate energy. 
The function $\mathcal{R}(\w;a,b;n,m)$ will in general depend not only on the frequency, but also the orientation of the spins $a,\,b$ and on their locations $n,m$. 
However, in the absence of long-range interactions causing correlations between spins at spatially-separate locations, or for short times, $\langle s^a_n(\w) s^b_m(\w) \rangle \propto \langle s^a_n(\w)\,s^b_n (\w)\rangle \delta_{nm}$. 
Therefore, we impose $\overline{S_{s^a_n s^b_m}}(\w) = \overline{S_{s^a_n s^b_n}}(\w)\,\delta_{nm}$. 
As a result, we can map Eq.~\eqref{apeq:DissRate} onto an equivalent expression given in terms of the thermally-averaged spin PSD $\overline{S_{s^a_n s^b_n}}(\w)$, without needing to explicitly compute the PSD.
We demonstrate this approach below.

We proceed by defining the non-thermally-averaged PSD for the spin operators as
\be
2 \pi \delta( \w - \w') \, S_{s_n^a s_m^b}(\w)  \equiv \frac{1}{2} \bra{\psi} s_n^a(\w) s_m^b( -\w') + s_m^b( -\w') s_n^a(\w) \ket{\psi},
\label{apeq:PSDDef}
\ee
where $\psi$ denotes the state of the system.
This definition has the properties required to be mapped onto a classical PSD.
Setting the state of the system $\ket{\psi}$ to be a thermal state we read off the PSD as
\be
\overline{S_{s_n^a s_m^b}}(\w)  = \pi \left(1 + e^{-\beta \w} \right) \sum_{k,l} \, \frac{e^{-\beta E_l}}{Z} \bra{l} s_n^a \ket{k} \bra{k} s_m^b \ket{l} \, \delta( \w - \w_{kl}),
\label{apeq:Soverline}
\ee
where $\w_{kl}$ is the energy difference between states $\ket{k}$ and $\ket{l}$ and $Z$ is the partition function.
We can compute $Q$ directly using Fermi's golden rule and see its dependence on the thermal PSD.
For the thermally averaged energy loss we find that $Q_{\QM}$, which is given semi-classically by Eq.~\eqref{apeq:DissRate}, is also given quantum-mechanically by
\be
Q_{\QM} =  \pi \gamma^2 \sum_{m,n} \sum_{k,l} \int \frac{d \w}{2 \pi} \,\w \,  B^a_n(\w) B^b_m(-\w) \left( 1- e^{-\beta \w} \right) \frac{e^{-\beta E_l}}{Z}\bra{l}s^a_n \ket{k} \bra{k}s_m^b\ket{l}\, \delta( \w - \w_{kl}).
\label{apeq:DissRateQM}
\ee
This is precisely of the form given in Eq.~\eqref{apeq:RoughQ}, with $\mathcal{R}(\w;a,b;n,m)$ given by
\be
\mathcal{R}(\w;a,b;n,m) = \gamma^2 \sum_{n,m} \,\w \, B^a_{n}(\w)\, B^{b}_{m}(-\w) \tanh\!\left[ \frac{\beta \w}{2}\right] \ .
\ee
If the semi-classical and quantum mechanical rates of energy loss are identified -- specifically, we equate Eqs.~\eqref{apeq:DissRate} and \eqref{apeq:DissRateQM} -- then using Eq.~\eqref{apeq:Soverline} we can extract the spin PSD as
\be
\gamma^2 \, \overline{S_{s^a s^b}} =  \frac{i}{2} \left[ \alpha_{ab}^* - \alpha_{ba} \right] \coth\! \left[ \frac{\beta \w}{2} \right] \!.
\ee
However, the quantity that we are really interested in is the PSD of the magnetisation vector.
As in Eq.~\eqref{apeq:MagnetisationVector}, we can take a volume average over the entire sample using the appropriate smearing function, obtaining the following form for the magnetisation vector in terms of the average spin state
\be
M^a \left( t \right) = \frac{\gamma}{V} \sum_n \langle s_n^a \left(t \right) \rangle \, .
\ee
This object is of primary interest in an experiment when length scales smaller than the dimensions of the sample are unresolvable.
We see that this definition and the definitions of the microscopic and macroscopic susceptibilities imply the identification
\be
\chi_{ab}(\w) = \frac{1}{V} \sum_n \alpha_{ab}(\w).
\ee
Using this crude smearing function results in the PSD
\be
S_{ab} \equiv S_{M^aM^b} = \frac{i}{2 V} \left[ \chi_{ab}^* - \chi_{ba} \right] \coth \!\left[ \frac{\beta \w}{2} \right]\!, 
\ee
which (as stated above) should be used on scales where individual spins are completely unresolvable.
Changing the smearing function such that there is manifest position dependence results in a different coefficient which could be computed or determined through calibration.
In identifying the PSDs we have implicitly assumed that the spins are not correlated with each other.
Note that in using Bloch's equations of the form given in Eq.~\eqref{apeq:BlochEqs}, we have effectively ignored any spatial dependence in $\bM$, so we are already implicitly assuming that we can ignore this information.

Using the causality properties enforces the conditions $\chi_{xx} = \chi_{yy}$ and $\chi_{xy} = - \chi_{yx}$, which themselves imply
\be
S_{xx} = \frac{1}{V} \, \text{Im} \left( \chi_{xx} \right) \, \coth\!\left[ \frac{\beta \w}{2} \right] =S_{yy},
\hspace{1cm}
S_{xy} = \frac{i}{V} \, \text{Re} \left( \chi_{xy} \right) \, \coth\!\left[ \frac{\beta \w}{2} \right] = - S_{yx} \, .
\ee
In the main text the problem was studied not in the Cartesian basis but in the $\pm$ basis.
We can also convert these results to that basis by noting that $\chi_{++} = \chi_{xx} - i \chi_{xy}$ and $\chi_{--} = \chi_{xx} + i \chi_{xy}$.
We then see
\be
S_{++} = S_{--} = 0,\hspace{1cm}
S_{-+} = \frac{1}{V} \coth\!\left[ \frac{\beta \w}{2} \right] \text{Im} \left( \chi_{++} \right)\!,\hspace{1cm}
S_{+-} = \frac{1}{V} \coth\!\left[ \frac{\beta \w}{2} \right] \text{Im} \left( \chi_{--} \right)\!.
\ee
To recover Eq.~\eqref{eq:SPNoise} as stated in the main text, we must take the high spin-temperature limit and further we have the following form of the imaginary part of the susceptibility,
\be
\text{Im} \left( \chi_{++} \right) = \frac{ M_0 T_2 \gamma}{1 + \left(\w - \w_0\right)^2 T_2^2}.
\ee
Note that the expression derived here differs from that of App.~SII in Ref.~\cite{Dror:2022xpi} by a $J$-dependent factor.
The correction arises from a more careful treatment here of taking the macroscopic limit.

\end{document}